\documentclass[
reprint,
superscriptaddress,
showpacs,
preprintnumbers,
nofootinbib,
nobibnotes,
amsmath,
amssymb, 
aps,
prd,
floatfix
]{revtex4-1}
\pdfoutput=1 
\usepackage[utf8]{inputenc} 
\usepackage{soul}
\usepackage{graphicx}  
\usepackage{dcolumn}   
\usepackage{bm,relsize}        
\usepackage{amssymb,amsmath,mathrsfs}
\usepackage{textcomp}
\usepackage{wasysym}
\usepackage{slashed}
\usepackage{multirow}
\usepackage{lipsum, color}
\usepackage[colorlinks=true,allcolors=purple]{hyperref}
\usepackage[usenames,dvipsnames,svgnames]{xcolor}
\usepackage{booktabs}
\usepackage{xfrac}
\usepackage[capitalize]{cleveref}
\usepackage{isotope}
\usepackage{accents}
\usepackage{comment}

\def\ltap{\ \raise.3ex\hbox{$<$\kern-.75em\lower1ex\hbox{$\sim$}}\ }
\def\gtap{\ \raise.3ex\hbox{$>$\kern-.75em\lower1ex\hbox{$\sim$}}\ }
\def\lsim{\ \raise.3ex\hbox{$<$\kern-.75em\lower1ex\hbox{$\sim$}}\ }
\def\gsim{\ \raise.3ex\hbox{$>$\kern-.75em\lower1ex\hbox{$\sim$}}\ }
\def\beq{\begin{equation}}
\def\eeq{\end{equation}}
 \def\be{\begin{equation}} \def\ee{\end{equation}}
\def\bea{\begin{eqnarray}} \def\eea{\end{eqnarray}}

\newcommand{\GeV}{{\, \rm GeV}}
\newcommand{\MeV}{{\, \rm MeV}}

\usepackage{fontawesome5} 
\definecolor{blue-violet}{rgb}{0.33, 0.17, 0.89}

\begin{document}

\title{Muon-induced baryon number violation}

\author{Patrick J.~Fox}
\email{pjfox@fnal.gov}
\affiliation{Theory Division, Fermilab, Batavia, IL 60510, USA}
\author{Matheus Hostert}
\email{mhostert@g.harvard.edu}
\affiliation{Department of Physics \& Laboratory for Particle Physics and Cosmology, Harvard University, Cambridge, MA 02138, USA}
\author{Tony Menzo}
\email{menzoad@mail.uc.edu}
\affiliation{Department of Physics, University of Cincinnati, Cincinnati, Ohio 45221, USA}
\author{Maxim Pospelov}
\email{pospelov@umn.edu}
\affiliation{School of Physics and Astronomy, University of Minnesota, Minneapolis, MN 55455, USA}
\affiliation{William I. Fine Theoretical Physics Institute, School of Physics and Astronomy, University of Minnesota, Minneapolis, MN 55455, USA}
\author{Jure Zupan}
\email{zupanje@ucmail.uc.edu}
\affiliation{Department of Physics, University of Cincinnati, Cincinnati, Ohio 45221, USA}

\preprint{FERMILAB-PUB-24-0314-T}

\date{\today}

\begin{abstract}
The search for charged-lepton flavor violation in muon capture on nuclei is a powerful probe of heavy new physics. 
A smoking gun signal for $\mu \to e$ conversion is a monochromatic electron with energy almost equal to the muon mass. 
We show that light new physics can mimic this signature and that it can also lead to electrons above the $\mu \to e$ signal peak. 
A concrete example of such light new physics is $\mu^- $-nucleon annihilation into a light dark sector, which can produce an energetic $e^-$ as well as $e^+e^-$ byproducts.
Due to the size of the muon mass, the exotic muon capture process can be kinematically allowed, while the otherwise stringent constraints, e.g., from proton decay, are kinematically forbidden.
We also discuss other relevant constraints, including those from the stability of nuclei and muon capture in the interior of neutron stars.
\end{abstract}

\maketitle

\section{Introduction} 

In the Standard Model (SM) of particle physics, taus, muons, and electrons do not change flavor.
With the addition of neutrino masses, the rate for conversion is non-zero but remains experimentally unobservable.
Therefore, charged-lepton flavor violation (cLFV) provides a smoking gun signature for new physics.
The most sensitive probes of cLFV include flavor-changing muon decays, such as $\mu \to e \gamma$ and $\mu \to eee$~\cite{Feinberg:1958zzb}, and muon-to-electron conversion on nuclei, $\mu^- A \to e^- A^*$~\cite{Marciano:1977cj}.
Due to significantly improved beam intensities and background reductions, next-generation experimental facilities are expected to achieve major leaps in sensitivity to these rare processes.
The Mu2e experiment at Fermilab~\cite{Mu2e:2012eea,Mu2e:2014fns,Mu2e-II:2022blh,CGroup:2022tli} and COMET~\cite{Kuno:2013mha,COMET:2018auw} experiment at J-PARC will reach $\mu \to e$ conversion rates over two orders of magnitude below current bounds by the end of the decade, while the DeeMe experiment at J-PARC~\cite{Teshima:2019orf} will improve bounds by an order of magnitude on a shorter timescale.
In terms of new-physics reach, these experiments will probe cLFV new physics at scales as high as $\Lambda \sim 10^8$~GeV (see Ref.~\cite{Fernandez-Martinez:2024bxg} for a recent global analysis of cLFV in effective theories).

It is also conceivable that light new particles can manifest in the ultra-rare rates probed by the cLFV experiments.
Some examples already studied in the literature include the production of an invisible axion-like-particle $X$ in muon decays such as the two body $\mu \to e X$ decays~\cite{GarciaiTormo:2011cyr,Uesaka:2020okd,Calibbi:2020jvd,Hill:2023dym,Knapen:2023zgi}, virtual exchange of axion-like-particles~\cite{Fuyuto:2023ogq}, as well as other signatures from muon decay induced production of light new particles that then further decay to visible SM particles --- electrons, positrons and/or photons~\cite{Echenard:2014lma,Heeck:2017xmg,Hostert:2023gpk,Knapen:2023iwg}. Note that in all these cases the kinematic reach for the production of new particles is set by the available energy, the mass difference $m_\mu-m_e$. 

In this manuscript, we explore a distinct new possibility: a class of models where the production of new particles is a result of $\mu^-$ annihilation with a proton or neutron bound inside a nucleus. The energy available for the production of new particles is now much larger, $\sim m_\mu + m_p$. 
However, the process also violates the baryon number and faces stringent constraints. 

These scenarios constitute an example of new physics models that can appear only in muon conversion experiments and would not be probed directly by other cLFV probes. 
As we will show below, the sample models we will discuss can be probed with the primary running mode and design parameters of Mu2e and COMET, complementing searches for light new physics that would otherwise rely on dedicated runs or on decays of muons in orbit~\cite{GarciaiTormo:2011cyr,Uesaka:2020okd,Xing:2022rob,Hill:2023dym}.

In choosing the new physics benchmarks, we have two questions in mind that we want to explore quantitatively in specific examples: 
\begin{enumerate}
\item Can light new physics `fake' a conventional $\mu \to e $ conversion signal?
\item Can light new physics lead to electrons or positrons with energies greater than $m_\mu$?
\end{enumerate}
We will show that the answer to both questions is yes. 
With the goal of providing proof of existence, we study a dark sector model where baryon number violation is induced by muonic reactions.
Constraints from proton decay and nuclear stability can be avoided via a judicious choice of new particle spectra and flavor structures.
Ultimately, our study demonstrates that there are other new physics signatures beyond the mono-energetic conversion electrons that the $\mu\to e$ experiments can look for.

This article is organized as follows. 
In \cref{sec:capture} we describe muon-proton annihilation induced muon capture on nuclei and the resulting possible production channels of new particles.
We study quantitatively an example of such a transition in more detail in \cref{sec:darksectormodel}, deriving predictions for the Mu2e and COMET experiments and addressing constraints from other rare processes. \cref{sec:UVcompletions} discusses the necessary ingredients for the UV completion of these models, while \cref{sec:conclusions} contains our conclusions. 

\section{Particle production in muon capture}
\label{sec:capture}

\begin{figure}
    \centering
    \includegraphics[width=0.49\textwidth]{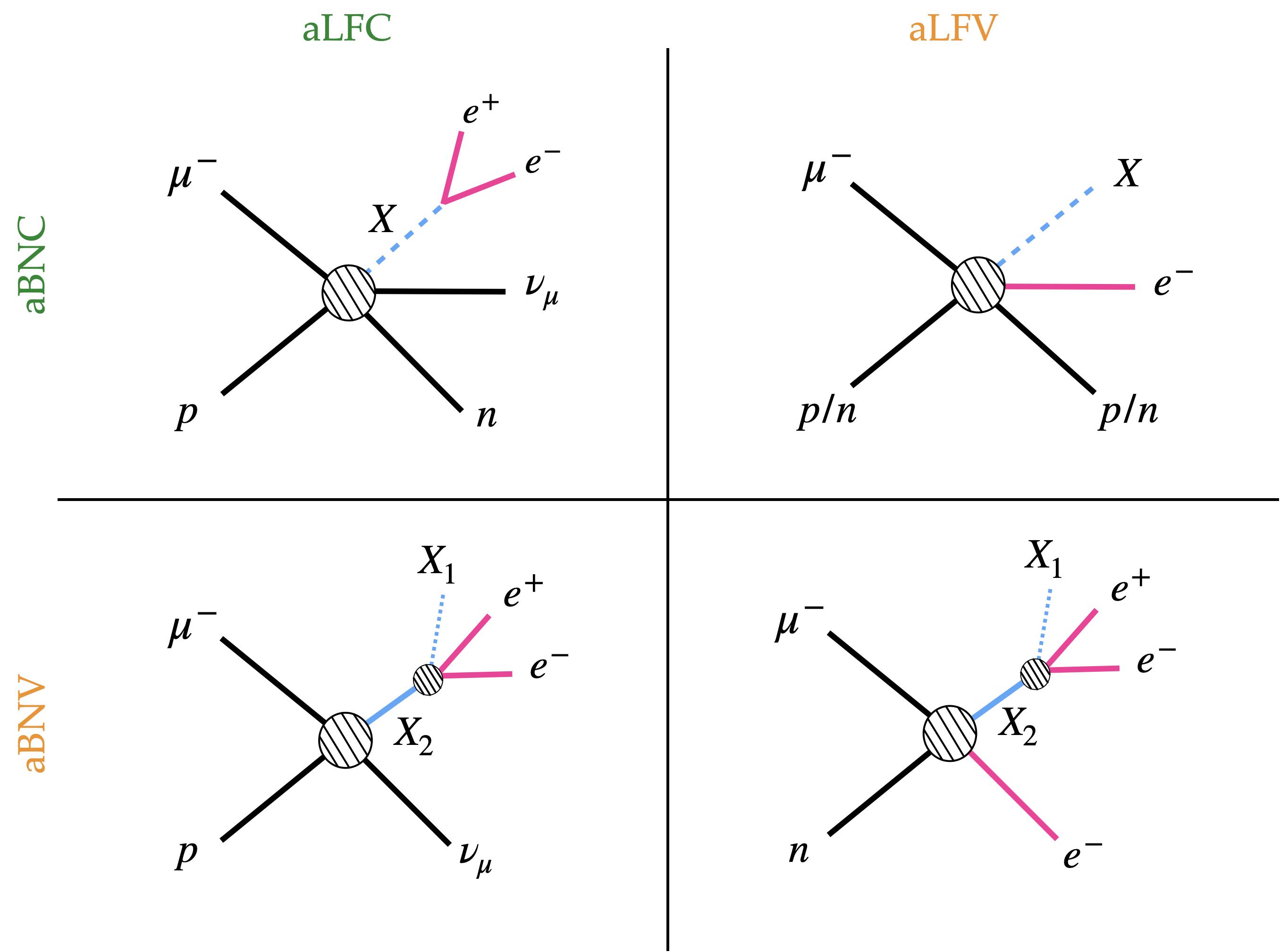}
    \caption{A schematic of the different particle production mechanisms in muon capture.
    Each quadrant corresponds to a different rule for baryon and lepton flavor number conservation or (apparent) violation.
    }
    \label{fig:enter-label}
\end{figure}

We are interested in models of light new physics that can generate the following types of transitions at the nucleon level,
\begin{align}\label{eq:general_capture}
\mu^- + p \to e^- + \{X\}^+,\text{~\,or~~}
\mu^- + n \to e^- + \{X\}^0,
\end{align}
where $\{X\}$ denotes dark and/or visible sector final state particles, out of which an odd number are fermions.
We are specifically interested in transitions with $e^-$ in the final state since these can result in a similar signal to the usual one in the $\mu \to e$ conversion searches but are still distinct from it when examined in detail.  
How much the process in \cref{eq:general_capture} will differ from the usual channel, $\mu^- + A \to e^- + A$, will depend on which particles are part of the $\{X\}^+$ and $\{X\}^0$ final states, as well as on the nature of the new interactions.
For instance, the electron may be a by-product of the decays in the dark sector, in which case the kinematics of these processes no longer ensure that the electron is mono-energetic. As is well appreciated in the literature, the fragmentation of on-shell ``dark" states $\{X\}$ opens up vast possibilities for the multiple production of light SM states, such as photons, electrons, and positrons. 

In $\mu \to e$ conversion, the small recoil energy of the nucleus can be neglected to a very good approximation. That is, the electron  takes away essentially all of the available energy, 
\beq
E_{e}^{\rm conv}\simeq m_\mu - E_b,
\eeq
where $E_b\sim (Z\alpha)^2 m_\mu/2 \sim  450$\,keV is the binding energy of the muonic aluminum atom, and thus $E_e^{\rm conv}\simeq 104.98$\, MeV~\cite{Haxton:2022piv}.
The distinct signature of $\mu \to e$ conversion is, therefore, mono-energetic electrons, with energy slightly below the muon mass and thus above the SM backgrounds.
The kinematics of $\mu\to e$ transitions that involve the production of light new physics, Eq.~\eqref{eq:general_capture}, is different. Depending on the details of the model, i.e., which particles are present in the final state and what their masses are, the energy of the final state electron may be well below or even well above the muon mass. 
If the outgoing electron is significantly less energetic than $E_{e}^{\rm conv}$, the backgrounds from the decays of muons in orbit, as well as from radiative decays followed by either Dalitz or Bethe-Heitler conversion, 
are expected to be prohibitively large. 
Furthermore, due to the small acceptance of the tracker for low-transverse momentum electrons ($p_T \gtrsim 90$ MeV at Mu2e~\cite{Mu2e:2014fns}), such less energetic electrons and other visible particles may be experimentally unobservable.
Therefore, we will be particularly interested in new physics scenarios in which $E_e > E_e^{\rm conv}$.

In what follows, we start with the general discussion of the minimal scenarios for the light particle production in muon capture, Eq.~\eqref{eq:general_capture}, and then focus on a particular concrete example, the model discussed in \cref{sec:darksectormodel}.
In general, we can divide the models into apparent baryon number conserving \textcolor{Green}{(aBNC)} and apparent baryon number violating \textcolor{Orange}{(aBNV)} scenarios, and also into apparent lepton flavor conserving \textcolor{Green}{(aLFC)} and apparent lepton flavor violating \textcolor{Orange}{(aLFV)} scenarios, giving four distinct classes of models. 
By ``apparent conservation or violation'', 
we refer to the conservation of baryon number or lepton flavors within visible sector particles, including neutrinos. 
These two quantum numbers need not necessarily be violated since the dark sector particles can also carry baryon or lepton numbers. 
Our discussion focuses on processes involving muonic aluminum atoms ($\mu\,\isotope[27\!]{Al}$), since $\isotope[27\!]{Al}$ is the currently planned target material in the Mu2e experiment. Should other targets for muon capture be used, our models can be easily generalized to those target atoms as well. 
Notably, the choice of a different target only changes $E_e^{\rm conv}$ by at most $\approx 5\%$, and thus has negligible impact on the kinematics of the final state particles for the models we consider~\cite{Haxton:2022piv}.  

\textbf{\textcolor{Green}{aLFC} and \textcolor{Green}{aBNC} scenarios}. These involve charged-current (CC) reactions, such as,
\begin{align}
    \mu^- + \isotope[27]{Al}&\to (X_0 \to \nu_\mu e^+e^-) +\isotope[27]{Mg}^{(*)},\label{eq:HNL}
    \\
    \mu^- + \isotope[27]{Al} &\to \nu_\mu + (X_0 \to e^+ e^-) +\isotope[27]{Mg}^{(*)}, \label{eq:vector}
\end{align}
where $X_{0}$ is an electromagnetically neutral dark sector particle. 
For example, $X_0$ in \cref{eq:HNL} could be a heavy neutral lepton, $X_0 = N$, produced through electroweak charge-current interaction, $\mu^- p \to N n$, that is induced via $N$ mixing with the SM neutrinos, followed by the $N\to \nu_\mu e^+e^-$ decay.  
The sterile-active neutrino mixing is strongly constrained for light sterile neutrinos, $m_N \lesssim m_\mu$. 
However, it could still induce the muon-to-$N$ capture rate as large as $R\sim 10^{-12} - 10^{-10}$ (see Eq.~\eqref{eq:R:def} below for the definition of $R$).
The sterile neutrino $N$ can decay within the target, for instance,  if the decay is mediated through a new dark force, such as $N \to \nu (A^\prime\to e^+e^-)$, where $A'$ is a dark photon.
Related NP transitions have been considered in the literature; for instance, the muon capture producing a sterile neutrino, $\mu^- p \to (N \to \nu \gamma) n$ was discussed in Ref.~\cite{McKeen:2010rx}.  The dark photon $A'$ can also be the dark state in the transition in \cref{eq:vector}. In the notation of \cref{eq:vector},  $X_0 = A^\prime$, so that the dark photon gets produced in the $\mu^- p \to \nu_\mu n A'$ transition and then decays to an electron-positron pair, $A'\to e^+e^-$.
If the $A^\prime$ boson is emitted directly, as in \cref{eq:vector}, it can be emitted from any of the lines in the diagram through couplings to muon, neutrinos, or nucleons.

In both types of transitions, Eqs.~\eqref{eq:HNL} and \eqref{eq:vector}, the energy of the outgoing electron is significantly smaller than $m_\mu$ (as is the energy of the outgoing positron). The kinematic end-point is given by the back-to-back decay configuration of electron recoiling against all the other particles, resulting in $E_e\lesssim m_\mu/2$.
This kinematic limitation is a result of both the fact that the above CC reactions conserve baryon number and thus the masses of final and initial nuclei are very close to each other, $M({\isotope[27]{Mg}}) - M({\isotope[27]{Al}}) \sim 3$~MeV, as well as due to the fact that the remaining final state particles, $e^+$ and $\nu_\mu$, are very light. 
Therefore, while the current constraints on the couplings of new particles do not preclude several thousands of such transitions from occurring at Mu2e or COMET, identifying the new signal events would be challenging for the current experimental designs.

\textbf{\textcolor{Orange}{aLFV} and \textcolor{Green}{aBNC} scenarios.} An example is
\begin{equation}
    \mu^- + \isotope[27]{Al}\to e^- + X_0 + \isotope[27]{Al},
\end{equation}
which could be a result of a $\mu^-\to e^- X_0$ decay, with nucleus not playing any significant role. The $X_0$ can be stable on collider time-scales, resulting in $\mu^- + \isotope[27]{Al}\to e^- + \isotope[27]{Al}+\text{inv}$ signature. A concrete realization of this scenario is an axion-like-particle, $X_0=a$, coupled to the $\overline{\mu}\gamma^5 e$ current \cite{GarciaiTormo:2011cyr,Uesaka:2020okd,Calibbi:2020jvd,Knapen:2023zgi,Hill:2023dym,Calibbi:2016hwq}. 
However, searching for $\mu^-\to e^-a$ decays is intrinsically more challenging than searching for the signal of $\mu \to e$ conversion, since in the former the electron is necessarily much softer, $E_e\lesssim m_\mu/2$. 
Similar conclusions apply also in the case the decays of $X_0$ are visible, such as $X_0 \to e^+e^-$.

\textbf{\textcolor{Green}{aLFC} and \textcolor{Orange}{aBNV} scenarios}. The simplest transitions of this type are
\begin{align}
    \mu^- + \isotope[27]{Al} &\to \nu_\mu + \left( X_1 \to e^+ e^- \right) + \isotope[26]{Mg}^{(*)},  
    \label{eq:BNV_proton_scalar}
    \\
    \mu^- + \isotope[27]{Al} &\to \nu_\mu + \left( X_2 \to X_1  e^+  e^-\right) + \isotope[26]{Mg}^{(*)},
    \label{eq:BNV_proton}
\end{align}
where $X_{1,2}$ are dark sector states. 
Compared to the previous two types of models, the main new ingredient is that the number of SM baryons changes. 
In principle, this releases an additional $m_N\sim 1\,$GeV of energy and thus one can have $E_e > E_e^{\rm conv}$, as long as $m_1 \lesssim m_{n},m_p$. 

Note that the transitions in \cref{eq:BNV_proton_scalar,eq:BNV_proton} need not be BNV since $X_{1,2}$ can carry nonzero baryon number. 
Nevertheless, in general, the constraints on BNV processes, i.e., the bounds on the (in)stability of the proton and the bound neutrons, do place stringent bounds on aBNV processes. 
That is, the transitions in \cref{eq:BNV_proton_scalar,eq:BNV_proton} above imply nucleon decays either through tree-level diagrams or via loop diagrams with off-shell muons, which then naively precludes any signatures at the $\mu \to e$ conversion experiments. 
However, as we will show below, through a combination of judicious charge assignments and kinematic suppressions in the form of mass hierarchies, these contributions can be sufficiently suppressed either due to off-shellness, higher loop orders, or phase space suppression, ensuring phenomenologically-viable nucleon stability.  

\begin{figure}[t]
    \centering
    \includegraphics[width=0.49\textwidth]{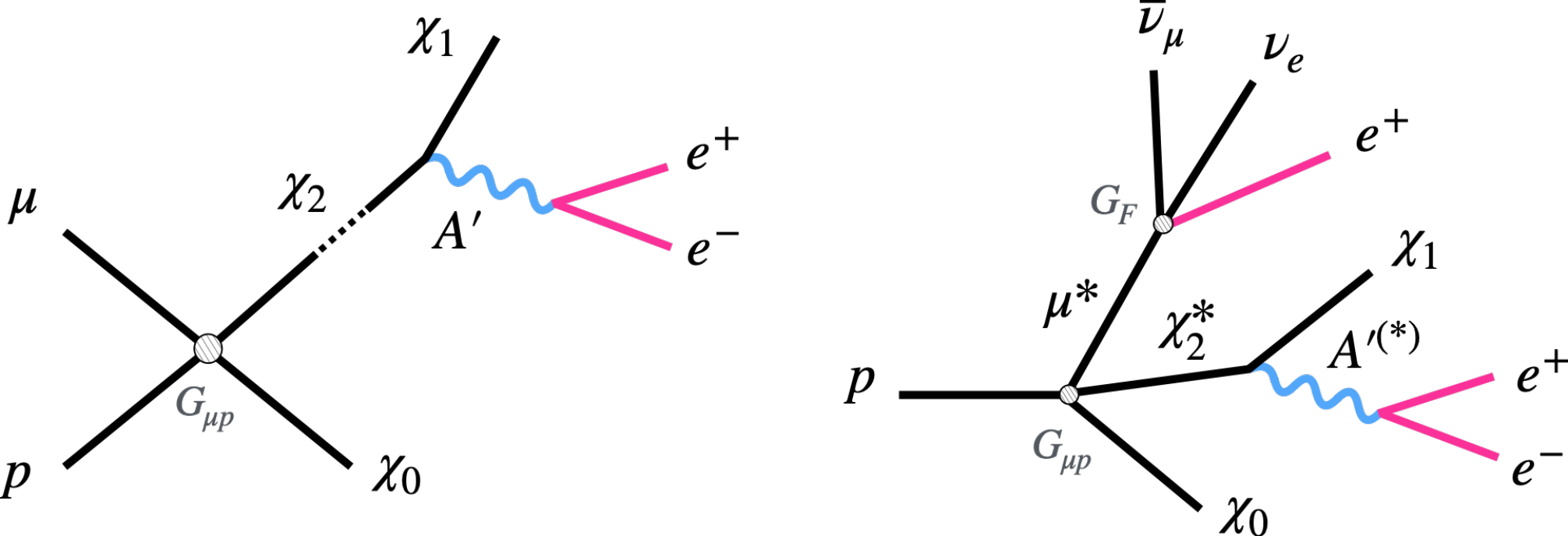}
    \caption{The scenario for muon-induced (apparent) baryon number violation considered in this work (\cref{eq:modelI}). 
    On the left, we show the muon capture on protons, and on the right, the tree-level proton decay channel.
    In both cases, the dark photon $A^\prime$ can be either on or off shell.
    \label{fig:diagrams}}
\end{figure}

In the two examples above, the nucleon decay transitions are  $p\to  \mu^{+*} \nu_\mu X_{1,2}$. These are kinematically forbidden if the mass spectra satisfy $m_{X_{1,2}} > m_p - m_e$, neglecting neutrino masses and using the fact that the off-shell $\mu^{+*}$ decays leptonically, $\mu^+ \to e^+ \overline{\nu}_\mu \nu_e$. 
Since $X_1$ and $X_2$ are allowed to decay, one also needs to consider the following full decay chains, 
\begin{align}
    p&\to \nu_\mu + (X_1^*\to e^+ e^-) + (\mu^{+*} \to e^+  \nu_e  \bar \nu_\mu),
    \\
    p&\to \nu_\mu + (X_2^* \to X_1  e^+ e^-) + (\mu^{+*} \to e^+ \nu_e \bar \nu_\mu),
\end{align}
as well as their variants. 
As we will show below, these decays can be sufficiently suppressed in the region of parameter space where $E_{e^+/e^-} > E_e^{\rm conv}$.
We will illustrate this within the scenario in \cref{eq:BNV_proton}, identifying  $X_{1,2}$ with dark sector fermions, $X_{1,2} = \chi_{1,2}$, and also generalizing it by replacing $\nu_\mu$ with some generic light particle $\chi_0$.
The muon capture and the associated nucleon decay channel schematics for this case are shown in \cref{fig:diagrams}.

\textbf{\textcolor{Orange}{aLFV} and \textcolor{Orange}{aBNV} scenarios}. 
Transitions of this type are
\begin{align}
    \mu^- + \isotope[27]{Al} &\to e^- + X_1 + \isotope[26]{Al}, \label{eq:BNV_neutron_fail} 
    \\
    \mu^- + \isotope[27]{Al} &\to e^- + (X_2 \to X_1 e^+ e^-) + \isotope[26]{Al}, \label{eq:BNV_neutron} 
\end{align}
and are induced by the $\mu^-n \to e^- X$ nucleon level transitions. 
Similarly to other cases, these transitions can also result in energetic electrons above the $\mu\to e$ endpoint, $E_{e^-} \gtrsim E_e^{\rm conv}$.  
In \eqref{eq:BNV_neutron_fail}, this occurs for $m_{X_1} \lesssim m_n$, and in \cref{eq:BNV_neutron}, if either $m_{X_2} \lesssim m_n$ and/or if  $m_{X_2}\gg m_{X_1}$.

As in the previous classes of models, also here the most stringent constraints are due to the bounds on BNV nucleon decays.
For instance, if the decays of bound neutrons, $n \to \mu^{+(*)} e^- X_0$, are kinematically forbidden, this also means that the transition  \cref{eq:BNV_neutron_fail} will not result in 
energetic electrons in Mu2e. The decay chain in \cref{eq:BNV_neutron} is of greater phenomenological interest. 
Choosing $m_{X_1} > m_n$, the decay $n \to \mu^{+(*)} e^- (X_1^* \to X_0 e^+e^-)$ comes with a second virtual particle in the final state, which, combined with the phase-space suppression caused by the smallness of $\Delta E = m_n - m_{X_0} - 3m_e$, can make the decays of neutrons bound inside nuclei sufficiently rare, at least at tree-level.
In the transition in \cref{eq:BNV_neutron}, the electron from the primary LFV vertex is soft, $E_{e^-} < m_\mu$.
However, the positron and the electron from the $X_2 \to X_1 e^+e^-$ decay can be energetic enough to be observed in the experiment.

\section{Muon-induced baryon number violation}
\label{sec:darksectormodel}

\begin{figure*}[t]
    \centering
    \includegraphics[width=0.49\textwidth]{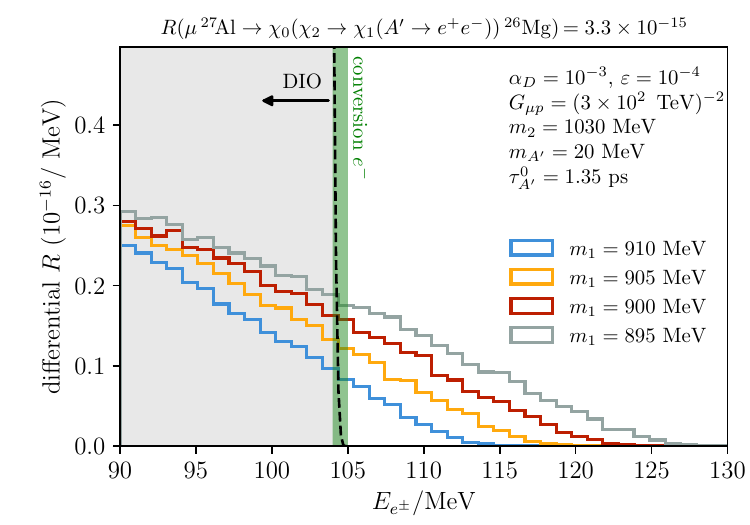}    \includegraphics[width=0.49\textwidth]{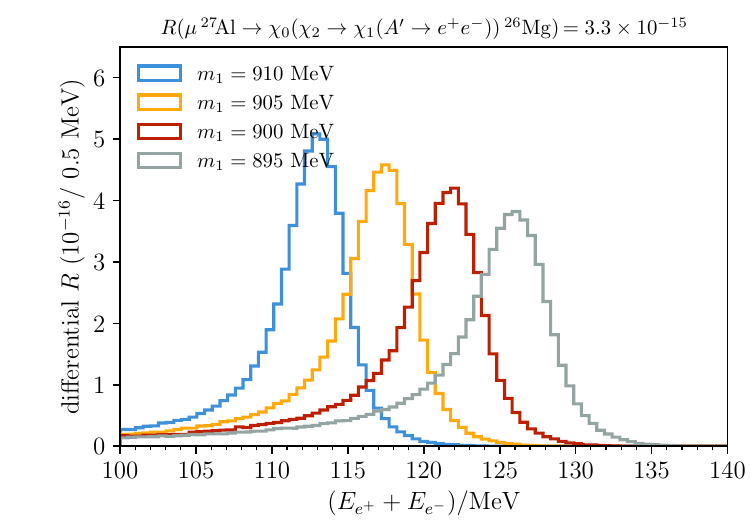}
    \\
    \includegraphics[width=0.49\textwidth]{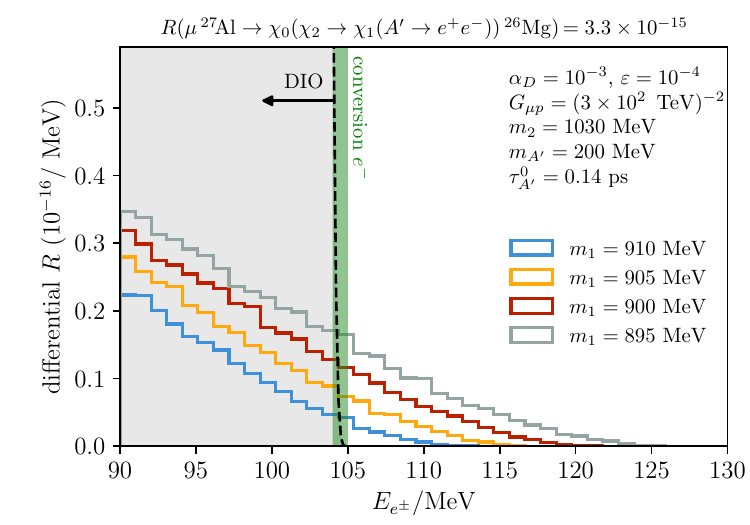}    \includegraphics[width=0.49\textwidth]{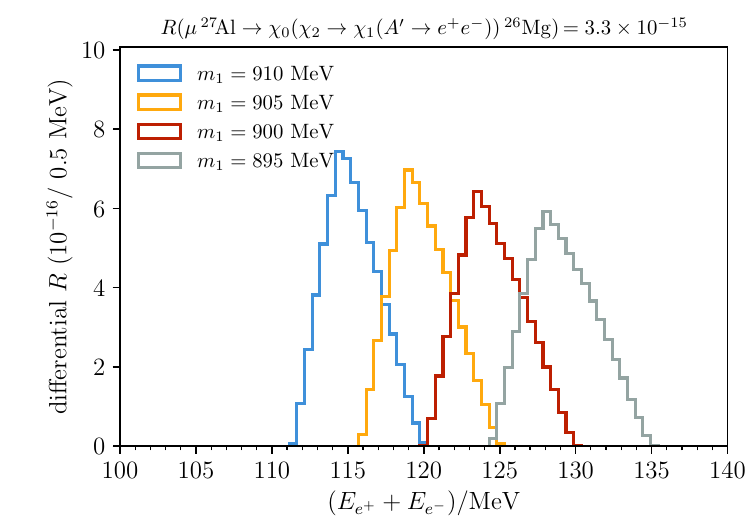}
    \caption{
    The energy spectrum for electrons and positrons produced in exotic muon capture for a few choices of dark particle masses.
    Couplings are fixed as indicated in the left panel of each row.
    On the left, we show the individual electron/positron energies, and on the right,  their combined energy.
    The usual $\mu \to e$ conversion electron energy is shown as a narrow green band around $E_{e^-}^{\rm cap} \simeq 104.98$~MeV, and the region where muon decay-in-orbit dominates is shaded in gray~\cite{Czarnecki:2011mx}.
    \label{fig:muon_capture_spectra}
    }
\end{figure*}

\begin{figure*}[t]
    \centering
    \includegraphics[width=0.49\textwidth]{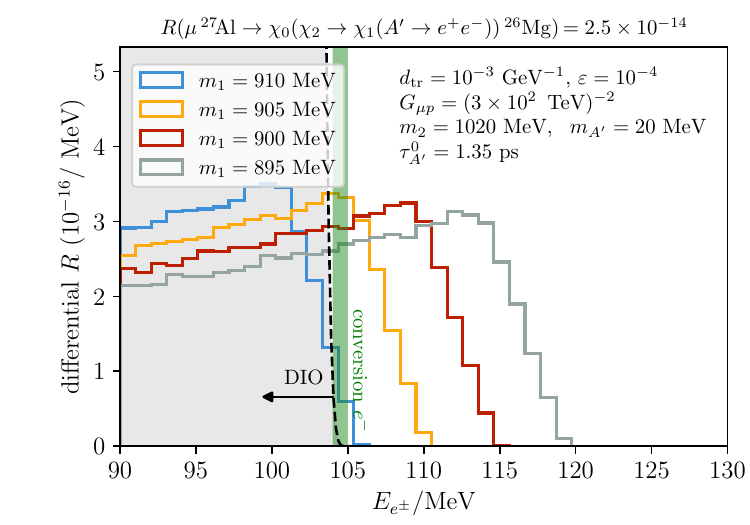}    
    \includegraphics[width=0.49\textwidth]{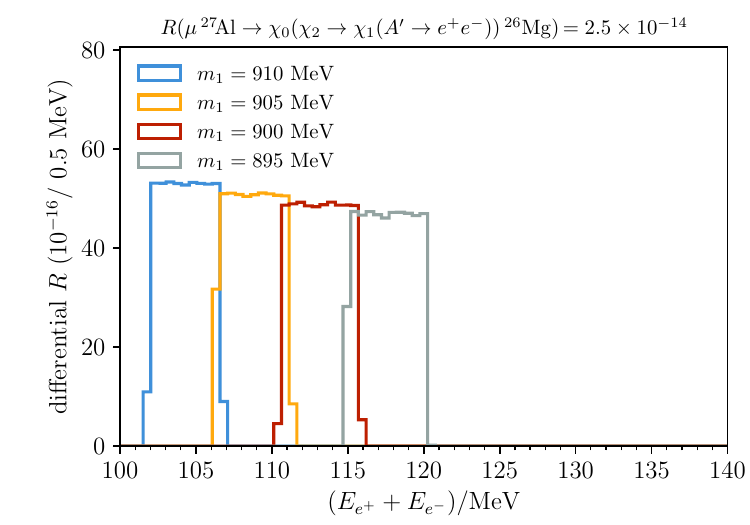}
    \\
    \includegraphics[width=0.49\textwidth]{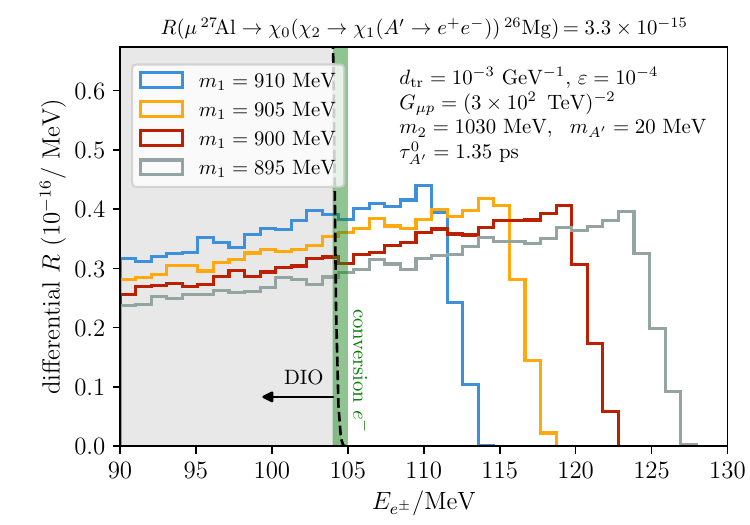}    
    \includegraphics[width=0.49\textwidth]{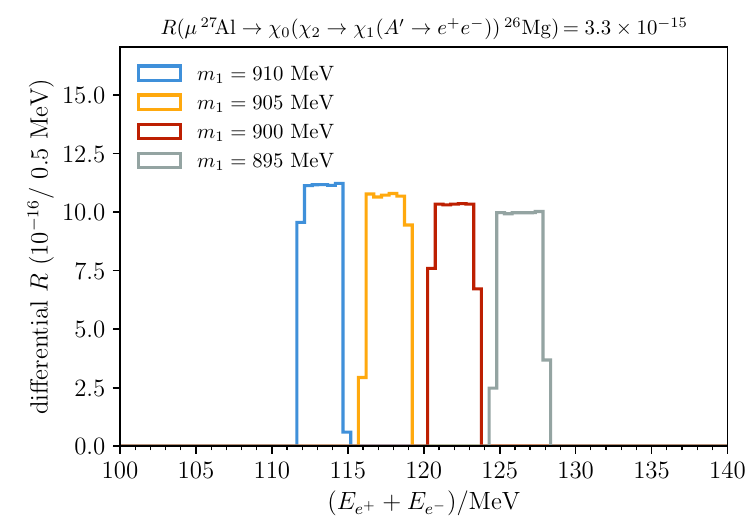}
    \caption{
    Same as \cref{fig:muon_capture_spectra} but for $\alpha_D \ll 1$ and $d_{\rm tr} = 10^{-3}$~GeV$^{-1}$. 
    The $\chi_2 \to \chi_1 e^+e^-$ transition is induced by the tensor interaction of \cref{eq:A':interaction} and leads to a preference for more energy-asymmetric $e^+e^-$.
    \label{fig:muon_capture_spectra_dtr}}
\end{figure*}

Next, we discuss in detail a concrete model of muon capture on a nucleus that results in multiple particle production. We specialize to the aLFC/aBNV scenario in Eq.\eqref{eq:BNV_proton}, but with three dark sector particles, $\chi_{0,1,2}$,
\beq
  \mu^- + \isotope[27]{Al} \to \chi_0 + \left( \chi_2 \to \chi_1  e^+  e^-\right) + \isotope[26]{Mg}^{(*)}.
    \label{eq:BNV_proton:extend}
\eeq
The transition conserves lepton flavor number if $\chi_0$ carries a muon lepton number (for instance, $\chi_0$ can be $\nu_\mu$ in which case this is exactly the scenario in Eq.~\eqref{eq:BNV_proton}). Similarly, the transition conserves baryon number if either $\chi_0$ or $\chi_{2,1}$ carry baryon number. 
We are particularly interested in the part of the parameter space that results in $e^+e^-$ pairs with high enough energy to be observable at Mu2e and COMET.

We start by realizing the scenario within a U$(1)_{\rm D}$ extension of the SM, to which we add three dark fermions, $\chi_{0,1,2}$ with masses $m_{0,1,2}$. 
The U$(1)_{\rm D}$ gauge symmetry is spontaneously broken, giving rise to a dark photon $A'$ with mass $m_{A'}$. Kinetic mixing gives rise to the interaction term
\beq \label{eq:A':mixing}
\mathscr{L} \supset  e \varepsilon A^\prime_\mu  J^\mu_{\rm EM},
\eeq
where $J^\mu_{\rm EM}$ is the SM electromagnetic current. 

The flavor-violating interaction at low energies is assumed to be given by the following effective dimension 6 operator,
\beq
\label{eq:modelI}
\mathscr{L} \supset  \frac{G_{\mu p}}{\sqrt{2}}  \left(\bar p \chi_2 \right) \left( \bar \mu \chi_0 \right) + \text{h.c.}
\eeq
One can also contemplate other chiral and Lorentz structures than the one above; we discuss some variations and their impact on our conclusions in \cref{sec:UVcompletions}.
At the quark level, the above interaction is generated from operators of dimension 9 or higher. 
The underlying new-physics scale associated with the nucleon-level four-fermion operator is, therefore, more accurately estimated as $\Lambda \sim \left( \Lambda_{\rm QCD}^3/G_{\mu p}\right)^{1/5}\sim 100$~GeV, which points to new color-charged particles at the electroweak scale (see \cref{sec:UVcompletions} for a more detailed discussion of UV completions).

Note that if $\chi_0$ carries baryon number, $B(\chi_0) = 1$, then baryon number is conserved by the interaction in \cref{eq:modelI}. Since we are interested in scenarios where $m_0 < m_p$, this means that $\chi_0$ would be the lightest baryon and remain stable. If we can consistently assign $L(\chi_0) = 1$, then lepton number is also conserved. The other possibility is that $B(\chi_2)=B(\chi_1)=1$, and $L(\chi_2)=L(\chi_1)=1$, which also keeps the baryon and lepton numbers conserved. In this case, the lightest baryon is $\chi_1$. Which of the dark sector particles carries baryon and lepton numbers has no bearing on the phenomenology that we are interested in, though. The important point for us is that there is a muon-induced exothermic transition between nucleons and dark sector particles that reduces the number of nucleons by one unit. 

The heaviest dark fermions are required to interact with dark photon via the off-diagonal terms, giving rise to $\chi_2 \to \chi_1 + A^\prime$ decays.
We consider two ways of generating such interactions, either through dimension 4 and dimension 5 operators,
\begin{equation}
\label{eq:A':interaction}
    \mathscr{L} \supset g_D \big(\bar \chi_2 \gamma^\mu \chi_1\big) A_\mu^\prime + d_{\rm tr} \big(\bar \chi_2 \sigma^{\mu\nu}\chi_1\big) F_{\mu \nu}^\prime + \text{h.c.}.
\end{equation}
The first term arises if $\chi_1$ and $\chi_2$ are maximally mixed Majorana fermions (a well-known example is, e.g.,  inelastic dark matter~\cite{Tucker-Smith:2001myb}).
The second term is a non-renormalizable transition magnetic moment between the two (Dirac or Majorana) fermions through the dark photon, where $F_{\mu\nu}^\prime = \partial_\mu A^\prime_\nu - \partial_\nu A^\prime_\mu$.
In the remainder of the paper, we will refer to the following benchmark for Mu2e signatures,
\beq
\begin{split}\label{eq:benchmarkI}
& \text{Benchmark (I)} 
\\
& m_2 = 1030\text{ MeV},\,m_1 = 900\text{ MeV},
\\
&m_0 = 0, \, m_{A^\prime} = 20\text{ MeV}, 
\\
&G_{\mu p} = (300 \text{ TeV})^{-2}, \, \varepsilon=10^{-4}, \, \alpha_D = 10^{-3}.
\end{split}
\eeq
Similarly, in the case of a transition magnetic moment,
\beq
\begin{split}\label{eq:benchmarkII}
& \text{Benchmark (II)} 
\\
& m_2 = 1030\text{ MeV},\,m_1 = 900\text{ MeV},
\\
&m_0 = 0, \, m_{A^\prime} = 20\text{ MeV}, 
\\
&G_{\mu p} = (300 \text{ TeV})^{-2}, \, \varepsilon=10^{-4}, \, d_{\rm tr} = 10^{-3} \text{ GeV}^{-1}.
\end{split}
\eeq
This choice of spectrum ensures that the energy release in $\mu \to e$ transitions can be large $E_{e^+/e^-} > E_e^{\rm conv}$ while guaranteeing that at least two particles must be off-shell in nucleon decay.

Given Eq.~(\ref{eq:A':mixing}) and our benchmarks, the partial width for the leptonic decay of the dark photon is given by
\begin{equation}
    \Gamma_{A' \rightarrow e^+ e^-} = \frac{1}{3} \alpha \varepsilon^2 m_{A'} \sqrt{1-4 r_e^2} (1 + 2 r_e^2)~,
\end{equation}
where $r_e \equiv m_e / m_{A'}$. 
For our benchmarks, this gives a lifetime of $\tau_{A^\prime}^0 \approx 1.4$ ps.
The $\chi_2$ lifetime depends on which interaction term in \cref{eq:A':interaction} is considered.  Assuming the channel is open, the partial width for $\chi_2$ to decay to $\chi_1$ and $A'$ through the vector interaction is given by
\begin{equation}
    \Gamma^V_{\chi_2 \rightarrow \chi_1 A'} = \frac{1}{4} \alpha_D \frac{m_2^3}{m_{A'}^2} F\left( r_1, r_{A'} \right)~,
\end{equation}
where
\beq
\begin{split}
    F(r_1, r_{A'}) =& \left(r_1^4 - 2r_{A'}^4 + r_1^2 (r_{A'}^2 - 2) + r_{A'}^2 (1 - 6 r_1) +1\right) \\
                    & \times \sqrt{r_1^4 - 2 r_1^2 (r_{A'}^2 + 1) + (1 - r_{A'}^2)^2}~,
\end{split}
\eeq
and $r_1 \equiv m_1 / m_2$ and $r_{A'} = m_{A'} / m_2$.  Giving a benchmark lifetime $\tau_{\chi_2}^0 \approx 6\times 10^{-14} \text{ ns}$.  
In the limit that $r_{A'}\rightarrow 0$, the mass splitting between $\chi_1$ and $\chi_2$ also vanishes ($r_1\rightarrow 0)$ and the width is well defined.
The tensor interaction gives the partial rate
\begin{equation}
    \Gamma^T_{\chi_2 \rightarrow \chi_1 A'} = \frac{1}{4\pi} d_{\text{tr}}^2 m_2^3 \tilde{F}(r_1, r_{A'})~,
\end{equation}
where
\begin{equation}
\begin{split}
    \tilde{F}(r_1, r_{A'}) =& \left(2(1-r_1^2)^2 - r_{A'}^2 (r_{A'}^2 + r_1^2 + 6 r_1 + 1)\right) \\
                            & \times \sqrt{r_1^4 - 2 r_1^2 (r_{A'}^2 + 1) + (1 - r_{A'}^2)^2},
\end{split}
\end{equation}
giving a benchmark lifetime of $\tau_{\chi_2}^0 \approx 3 \times 10^{-7} \text{ ns}$.
These decays are all well below the $\mathcal{O}(1\, \mu{\rm s})$ signal time window at Mu2e and COMET.
This is no longer true when the dark photon becomes off-shell in $\chi_2 \to \chi_1 e^+e^-$ decays. 
For instance, when $m_{A^\prime} = 200$~MeV, the $\chi_2$ lifetime is of the order of $\mu$s and $s$ for the couplings in \cref{eq:benchmarkI,eq:benchmarkII}, respectively. 
In comparison with the timescales of the beam and muon capture, these events will constitute a constant rate in time and may be harder to discriminate from beam-induced backgrounds.

\renewcommand{\arraystretch}{1.4}
\begin{table*}[t]
\centering
    \begin{tabular}{clll}
    \hline\hline
    No. &Decay channel & Mass range  & Decay rate scaling
    \\
        \hline
        \\[-0.4cm]
        1) & $p \rightarrow \mu \chi_2 \chi_0$ & $m_p > m_\mu +m_2   + m_0$ & \ \ \ $\Gamma \propto \dfrac{G_{\mu p}^2Q^5}{8\pi \times 16\pi^2} $
    \\
\multirow{ 2}{*}{2)} & \multirow{ 2}{*}{$p \rightarrow \mu (\chi_2^* \to \chi_1 A^\prime) \chi_0 \qquad\qquad\, \biggr\{$ } & $ m_p>m_\mu + m_{A^\prime}+ m_1 + m_0, $ & 
\multirow{ 2}{*}{\ \ \ $\Gamma \propto \dfrac{G_{\mu p}^2 \alpha_D Q^7}{(16\pi^2)^2 m_2^2}$}
   \\
   &  & $m_1+m_{A'}>m_2$ or $m_2>m_p-m_\mu-m_0$ & 
    \\
  \multirow{ 3}{*}{ 3)} & \multirow{ 3}{*}{$p \rightarrow \mu (\chi_2^* \to \chi_1 (A^\prime{}^*  \to ee)) \chi_0 \,\,\, \Biggr\{$} & $m_p > m_\mu  +2 m_e+ m_0 $ & 
  \multirow{ 2}{*}{\ \ \ $\Gamma \propto \dfrac{G_{\mu p}^2  \alpha_D (e\varepsilon)^2Q^{11}}{(16\pi^2)^3m_2^2 m_{A'}^4}$}
    \\
    & &  $m_1+2m_e>m_2$ or $m_2>m_p-m_\mu -m_0$ &
    \\
    & &  $2m_e >m_{A'} $ or $m_{A'}>m_p-m_\mu-m_0-m_1$ &
    \\
    4) & $p \rightarrow (\mu^* \to e \nu \nu) \chi_2 \chi_0$ & $m_p > m_2+m_0+m_e$ and $m_\mu> m_p-m_2-m_0$ & \ \ \ $\Gamma \propto \dfrac{G_{\mu p}^2 G_F^2 Q^{11}}{8\pi\times (16\pi^2)^3 m_\mu^2} $
    \\
 \multirow{ 3}{*}{  5)} &  \multirow{ 3}{*}{$p \rightarrow (\mu^* \to e \nu \nu) (\chi_2^* \to \chi_1 A^\prime) \chi_0\,\Biggr\{$} & $m_p > m_1+m_0+m_e+m_{A'}$  &  \multirow{ 3}{*}{\ \ \ $\Gamma \propto \dfrac{G_{\mu p}^2 G_F^2  \alpha_D Q^{13}}{(16\pi^2)^4m_\mu^2 m_2^2}$}
 \\
 & & $m_\mu> m_p-m_1-m_0-m_{A'}$ &
\\
 & &  $m_2 <m_1+m_{A'} $ or $m_p-m_0-m_e<m_2$ &
    \\
    \hline\hline
    \end{tabular}
    \caption{Proton decay channels and the corresponding constraints on the dark sector mass spectrum that we impose so as to forbid them at tree level. The naive scaling of each decay channel, assuming all final state particles are relativistic, is shown in the right-most column. Generalization to non-relativistic final state particles is straightforward -- see Sec.~\ref{sec:pdecay} for more details. \label{tab:proton_decay_channels}}
\end{table*}

\subsection{Muon capture signature}
\label{sec:mucapture_signal}

While the prediction for the aBNV transition $\mu^- + \isotope[27]{Al} \to \chi_0 + \left( \chi_2 \to \chi_1  e^+  e^-\right) + \isotope[26]{Mg}^{(*)}$ requires a calculation of inelastic nuclear response function, we can make a naive estimate for its rate by using the muon capture rates in the SM. That is, an ${\mathcal O}(1)$ fraction of muon capture on ${}^{27}$Al involves an emission of a neutron,  $\mu^- + \isotope[27]{Al} \to \nu_\mu +n + \isotope[26]{Mg}^{(*)}$, which has similar $1\to 3$ decay kinematics and involves the same types of nuclear wave-functions in the initial and final state. We assume that ignoring the phase space corrections, the ratio of aBNV and SM muon captures remains roughly the same, irrespective of the initial nucleus. This implies, for instance
\beq
\begin{split}
\frac{|\mathcal{M}(\mu \, \isotope[27]{Al} \to \chi_0 \chi_2 \,\isotope[26]{Mg})|^2}{|\mathcal{M}(\mu \, \isotope[27]{Al} \to \nu_\mu n \,\isotope[26]{Mg})|^2} \simeq &\frac{|\mathcal{M}(\mu p\to \chi_0 \chi_2)|^2}{|\mathcal{M}(\mu p\to \nu_\mu n)|^2} ~ .
\end{split}
\eeq
Including the phase space correction, we thus have
\beq
\Gamma(\mu \, \isotope[27]{Al} \to \chi_0 \chi_2 \,\isotope[26]{Mg})\simeq r_{\rm p.s.} \frac{G_{\mu p}^2}{G_F^2} \Gamma(\mu \, \isotope[27]{Al} \to \nu_\mu n \,\isotope[26]{Mg}),
\eeq
where we estimate the effect of phase space with the ratio of two Dalitz plot areas
\beq
\begin{split}
\label{eq:rps}
r_{\rm p.s.}\simeq &\frac{A_{\rm Dalitz}(\mu \, \isotope[27]{Al} \to \chi_0 \chi_2 \,\isotope[26]{Mg})}{A_{\rm Dalitz}(\mu \, \isotope[27]{Al} \to \nu_\mu n \,\isotope[26]{Mg})}
\\
\simeq & \frac{\big(M-m_{{\rm Mg}}-m_0-m_2\big)^2\big(M-m_{\rm Mg}+m_0+m_2\big)}{\big(M-m_{{\rm Mg}}-m_n\big)^2\big(M-m_{\rm Mg}+m_n\big)}.
\end{split}
\eeq
In the last line, we neglected the neutrino mass. 
Above, $M$ is the mass of the muonic atom, $M=m_\mu+m_{\rm Al}-E_b$, with $E_b\approx 0.463$\,MeV the binding energy, $m_{\rm Al}$ the mass of the $\isotope[27]{Al}$ and $m_{\rm Mg}$ the mass of the $\isotope[26]{Mg}$ atom. Note that for $m_0+m_2=m_n$ the above ratio is $r_{\rm p.s.}=1$. This is exact only if $m_0$ is massless, showing the limitation of our approximations.
For our benchmarks, \cref{eq:benchmarkI,eq:benchmarkII}, we find $r_{\rm p.s.} \simeq 4\times 10^{-3}$.

Equating $\Gamma(\mu \, \isotope[27]{Al} \to \nu_\mu n \,\isotope[26]{Mg})$ with the full SM muon capture rate, $ \Gamma_{\mu \text{Al}} = 0.7 \times \mu\mathrm{s}^{-1}= 4.6 \times 10^{-19}$~GeV~\cite{Suzuki:1987jf}, which is in line with our other approximations, we then get for the aBNV conversion rate 
\beq
\label{eq:R:def}
R \equiv \frac{\Gamma_{\text{exotic}}}{ \Gamma_{\mu \text{Al}}}\simeq  r_{\rm p.s.} \frac{G_{\mu p}^2}{G_F^2}.
\eeq
The benchmarks in \cref{eq:benchmarkI,eq:benchmarkII} result in a rate $R \sim 3\times 10^{-15}$, well within Mu2e and COMET's sensitivities and, as we will see, safe from nucleon decay bounds. 

\begin{figure*}[t]
    \centering
    \includegraphics[width=0.49\textwidth]{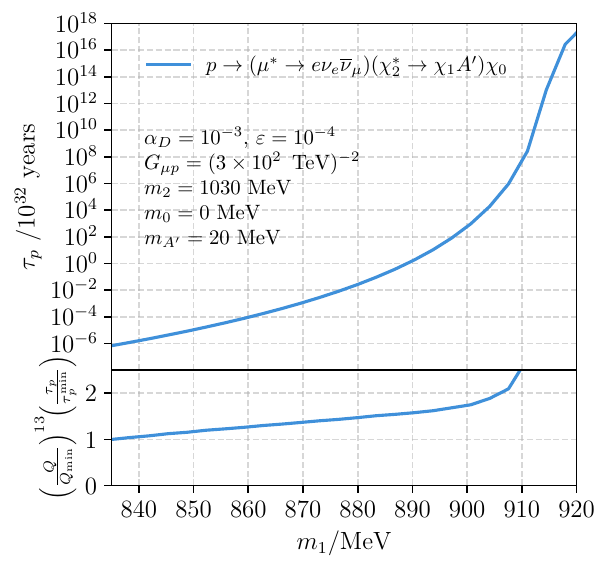}
    \includegraphics[width=0.49\textwidth]{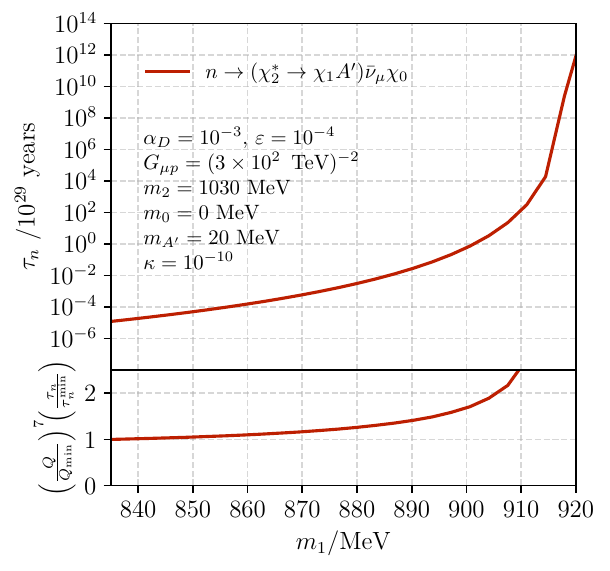}
    \caption{Estimates of the lifetime of a free proton (left) and neutron (right) as a function of the mass of the lightest dark particle $\chi_1$ for benchmark (I).
    The bottom panel illustrates the scaling of the lifetime with respect to the energy release $Q$, which is proportional to the nucleon mass minus the sum of the masses of all final state particles that are on shell, see \cref{eq:DeltaE,eq:Qdefn}. 
    The nucleon lifetime $\tau_N^{\rm min}$ and its respective energy release  $Q_{\rm min}$ are calculated for the smallest value of $\chi_1$ mass shown in the figure.
    We do not include loop-level proton decay channels which may dominate at the largest masses.
    \label{fig:plifetimes}
    }
\end{figure*}

Once produced, $\chi_2$ eventually decays to an $e^+e^-$ pairs through the cascade $\chi_2 \to \chi_1 (A^\prime \to e^+e^-)$.
The resulting energy spectra of the $e^+e^-$ pairs produced are shown for a few representative points in \cref{fig:muon_capture_spectra,fig:muon_capture_spectra_dtr}.
Note that the individual electron or positron energies can be greater than $m_\mu$ due to the additional energy released by the destruction of the proton mass.
This is also manifested in the sum of the electron and positron energies, which peaks at 
$E_{e^+} + E_{e^-} \simeq \left(m_2^2 - m_1^2 + m_{A^\prime}^2\right)/2m_2$, given that $\chi_2$ only has a small boost and thus the lab frame kinematics almost coincide with the rest frame $\chi_2\to \chi_1 A'$ two body decay kinematics ($E_{e^+}+E_{e^-}=E_{A'}$ for $A'$ on-shell).

Several features can help distinguish the aBNV processes from $\mu^- \to e^-$ conversion, most notably the differing electron spectra and the presence of positrons in the aBNV case.
A signal of equal magnitude with equal endpoints in the electron and positron channels that go beyond the $\mu^- \to e^-$ endpoint would be a strong indication for aBNV processes of the type considered here.
The positron signal is subject to even smaller backgrounds than the search for $\mu^- \to e^-$ signal is; the DIO background produces only electrons, and charge misidentification is expected to be rare.
Most background events are expected to arise from radiative muon capture (RMC) with either internal or external photon conversion. 
RMC, however, has an even lower endpoint than DIO, $E_{\rm RMC}^{\rm Al} \lesssim 101.9$~MeV~\cite{Yeo:2017fej}.
The positron search could be done already with the Al target, while one could also explore other targets (for a different BSM signal producing positrons, the lepton-number-violating $\mu^- \to e^+$ conversion, see Refs.~\cite{Yeo:2017fej,Lee:2021hnx}).
Finally, we note that the $Z$ dependence of the aBNV rate will differ from the one for $\mu^- \to e^-$ or $\mu^- \to e^+$ conversion (see, for instance, Ref.~\cite{Borrel:2024ylg}). 
The planned upgrades of Mu2e using, e.g., Au as the target~\cite{Mu2e-II:2022blh}, could thus help distinguish between these scenarios.

One of the crucial aspects of the Mu2e experiment is the requirement of a very high degree of initial proton and pion beam extinction within the signal time window, as well as a very reliable rejection of cosmic-ray-induced events. 
If a certain amount of pion ``contamination" is, for example, present within the signal-taking time interval, it may result in charged leptons with energy well above $E_e^{\rm conv}$. 
The $\pi^-$ capture on a nucleus can thus result in a $\sim 135$\,MeV photon, which can Compton-scatter and produce electrons well above $E_e^{\rm conv}$. 
The $\mu^-p$ annihilation electrons could, therefore, easily be misinterpreted as being due to a pion background. 
The above benchmarks for the $\mu^-p$ annihilation models, which can create $E_e \sim m_\pi$, thus show that extra care needs to be taken. In the event of a signal, a detailed investigation would be warranted to determine definitively whether such electrons do indeed stem from the pion decays and not from exotic new physics signatures, such as those investigated in this section.

\subsection{Proton decay}
\label{sec:pdecay}

\begin{figure*}[t]
    \centering
    \includegraphics[width=0.49\textwidth]{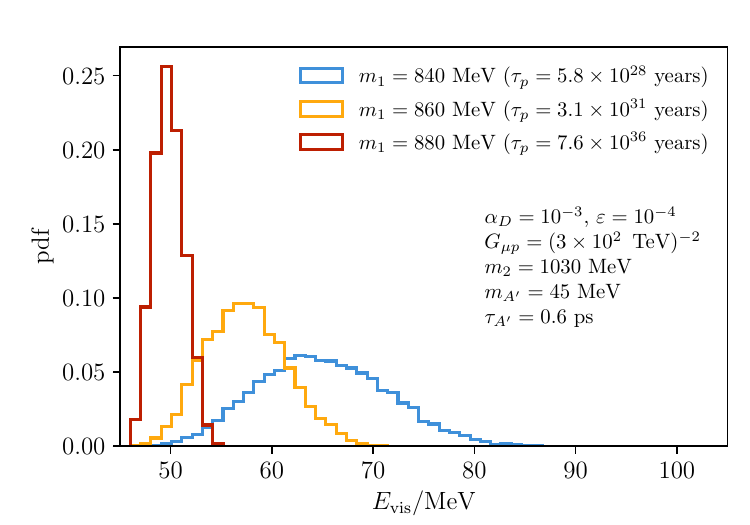}
    \includegraphics[width=0.49\textwidth]{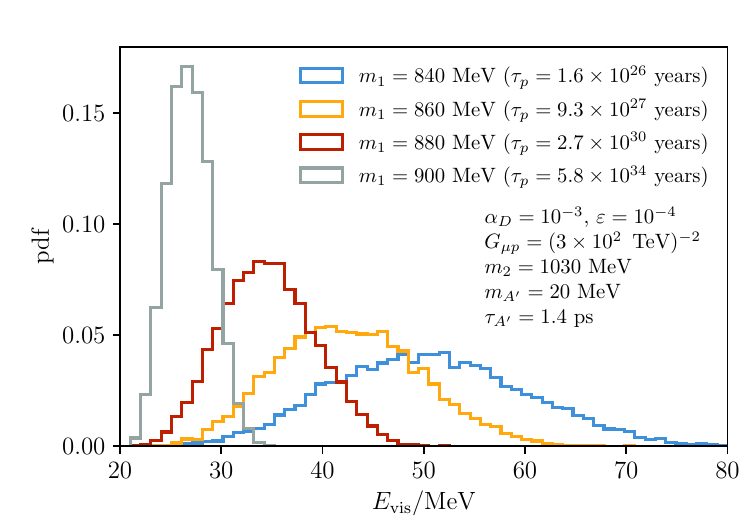}
    \caption{
    The normalized true total visible energy spectrum (energy of the $e^+e^+e^-$ system) in $p \to eee \nu \nu\chi_1 \chi_0$ decays for two choices of dark photon masses.
    The distributions look similar for the same benchmarks when using transition moments instead ($\alpha_D \to d_{\rm tr}$), although the total lifetimes increase.
    \label{fig:pdecay_spectra}}
\end{figure*}

The effective interaction in \cref{eq:modelI} violates proton number. 
What is an apparent phenomenological disaster can be readily reconciled with the most stringent constraints on proton decay by a judicious choice of mass hierarchies.  
By such a choice, we can ensure that some of $\chi_2, \chi_0, \mu$ must be produced off their mass shell, resulting in a higher particle multiplicity of final states and extra suppression by small coupling constants.   
Furthermore, the amount of open final state phase space will be small, suppressing the proton decay rate.
In \cref{tab:proton_decay_channels}, we list the possible decay channels, the necessary constraints on the mass hierarchy of the model to allow this channel, and an estimate for the proton decay width. 
Before discussing how these constraints affect the allowed region of parameter space, we first discuss how they are derived.  
Later on, we will calculate the rate fully numerically using \textsc{MadGraph5}\_{v3.5.3}.

\begin{figure*}[t]
    \centering
    \includegraphics[width=0.49\textwidth]{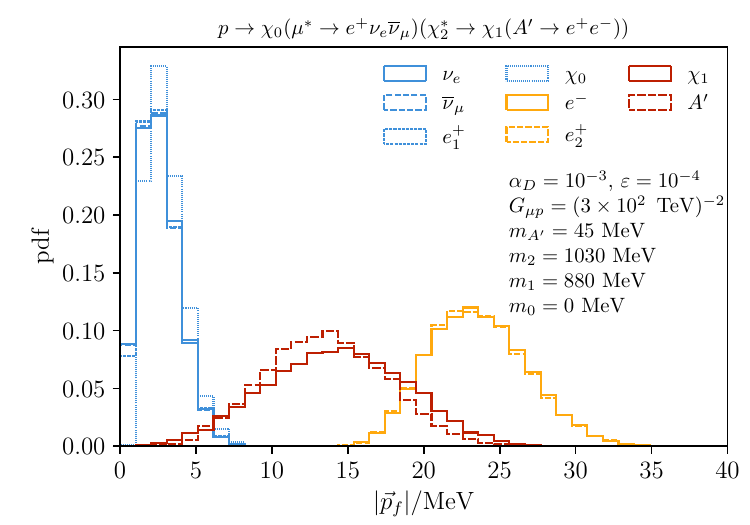}
    \includegraphics[width=0.49\textwidth]{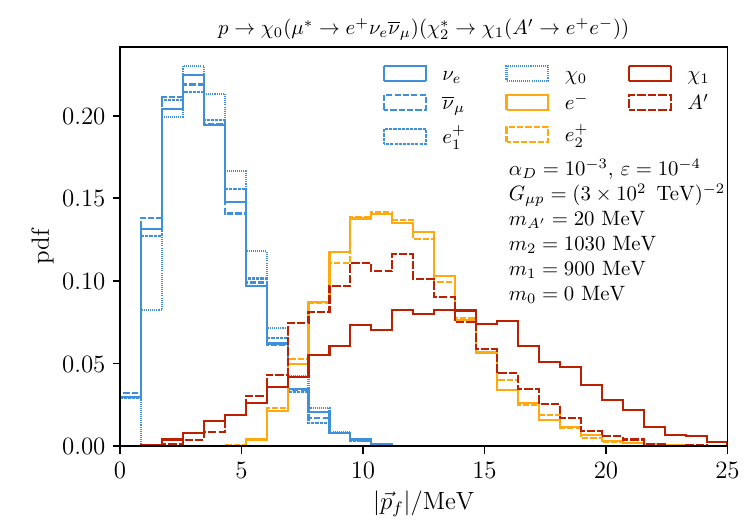}
    \caption{The momentum of the individual final state particles in 
 $p \to eee \nu \nu \chi_1 \chi_0$ decays for two choices of dark photon masses. Here the final state positrons $e^+_1$ and $e^+_2$ are distinguished at the event level by the magnitude of the three-momenta, with $e^+_2$ always being assigned to the larger momenta. 
\label{fig:pdecay_spectra_momenta}}
\end{figure*}

\paragraph{Scaling with energy release:} The energy released in proton decay is 
\beq\label{eq:DeltaE}
\Delta E=(m_p-\sum_i^{N_f} m_i),
\eeq 
where $N_f$ is the number of final state particles in the decay.
Each of the final particles therefore carries a typical momentum 
\beq\label{eq:Qdefn}
Q\simeq \Delta E/N_f~,
\eeq
with $N_f\in [3,7]$ depending on which decay is being considered; we leave the discussion of loop decay for later since it is UV model dependent.  
The typical value of $Q$ in proton decay is $\sim \mathcal{O}(5)$~MeV, see \cref{fig:pdecay_spectra_momenta}.
That is, at the upper range of $m_1$, the electrons will start becoming non-relativistic; we will discuss the effect of this below. In some of the decay channels, there are off-shell particles whose propagators we estimate to scale as inverse powers of the particle mass ($m^{-1}$ for fermions and $m^{-2}$ for bosons).  
For instance, the muon has 3 approximately massless decay products and scales as
\beq
\frac{\slashed{p}_\mu+m_\mu}{p_\mu^2-m_\mu^2}\sim \frac{3\slashed{Q}+m_\mu}{(3Q)^2-m_\mu^2}\sim -\frac{1}{m_\mu},
\eeq
whereas for $\chi_2$ the decay products may be non-relativistic and it scales as
\beq
\frac{\slashed{p}_{\chi_2}+m_2}{p_{\chi_2}^2-m_2^2}\sim \frac{\slashed{p}_{\chi_2}+m_2}{(m_{\chi_1}+m_{A'})^2-m_2^2}\sim -\frac{1}{m_2},
\eeq
where we used the fact that $\chi_1$ and $A'$ are non-relativistic (a better approximation for $\chi_1$ than it is for $A'$).  The matrix element squared acquires scalings with $Q$ when carrying out spin/polarization sums.  
The final state fermions contribute to $|\mathcal{M}|^2$ parametrically as $\sim Q$ $(\sim m)$ if they are relativistic (non-relativistic), while the final state vector bosons as $\sim Q^0$ $(\sim Q^2/m_{A'}^2)$.  
Furthermore, although we focus on the contact interaction, if the coupling is through the dipole interaction $d_{\rm tr}$ then $|\mathcal{M}|^2$ picks up another factor of $Q^2$.
The remaining $Q$ scaling comes from the phase space integrals, where for each final state particle
\beq
\frac{d^3 p_i}{(2\pi)^3 2E_i}\sim
\left\{
\begin{matrix}
\dfrac{1}{16 \pi^2} Q^2,& \text{relativistic},
\\[0.1in]
\dfrac{1}{16 \pi^2} \dfrac{Q^3}{m_i},& \text{non-rel.},
\end{matrix}
\right.
\eeq
The scaling of the matrix element and phase space for final state fermions is such that each one contributes $Q^3$ in both the relativistic and non-relativistic limits.
We present the results of these approximations in \cref{tab:proton_decay_channels}, where we have assumed the final state $A'$ is non-relativistic, which will be the case for the parameter space we are interested in.  
In \cref{tab:proton_decay_channels}, we also list the scaling for a tree level $p\to \mu ee \chi_1\chi_0$ decay with $\chi_1$ heavy enough for $A'$ to be required to be off-shell, which gives an additional $Q^4/m_{A'}^4$ suppression for the decay width. 

From these scalings, it is clear that channels 1 and 2 will lead to too rapid proton decay unless $\alpha_D$ is taken very small, which removes the prompt signal at Mu2e.  Channel 3 requires considerable tuning in the mass spectrum, so we ignore it from now on.  
This leaves channels 4 and 5, both of which can be small enough to avoid constraints.  
The benchmarks described in \cref{eq:benchmarkI,eq:benchmarkII} have a decay through channel (5) with an estimated lifetime of $\tau_p \sim 5\times10^{34}$ years and $\tau_p \sim 2\times10^{41}$ years, respectively.

\paragraph{MadGraph estimates:} To calculate the rate more accurately and study the resulting kinematics, we implement the model in \textsc{MadGraph5}\_{v3.5.3} and calculate the proton lifetime by generating $p \to (\mu^* \to e \nu_e \overline{\nu}_\mu)(\chi_2^* \to \chi_1 (A^\prime \to e\bar e)) \chi_0$ decays. 
This includes contributions from real as well as virtual $A^\prime$.
We calculate the decay width of the new particles for each choice of couplings. 
For values of $\alpha_D$ larger than about $\alpha_D = 10^{-2}$, the $\chi_2$ width can lead to important corrections to the decay rate and kinematics, but we do not explore this further here.
For the small values of kinetic mixing we are interested in, the width of the dark photon is always negligible. 
Our results are shown for $10^{5}$ generated events.

The left panel of \cref{fig:plifetimes} shows the resulting proton lifetime in the region of interest for the benchmark point in \cref{eq:benchmarkI} as a function of the $\chi_1$ mass.  
The bottom row of the left panel shows the accuracy of the $Q^{13}$ scaling.
This lifetime should be compared to experimental constraints.
The strongest constraints on the proton lifetime were obtained in underground neutrino detectors, where, for instance, $\tau(p \to e^+ \pi^0) >  2.4 \times 10^{34}$~years~\cite{Super-Kamiokande:2020wjk}. 
Other similar channels have been searched for, including $\tau(p \to e^+e^+e^-) >  3.4 \times 10^{34}$~years~\cite{Super-Kamiokande:2020tor}.
While these channels have similar topologies to $p \to e e e \nu \nu \chi_1\chi_0$, namely three electromagnetic showers, the kinematics differs significantly from our signal.
In particular, the energy of the three showers must reconstruct the proton, while in our case, the energy release is, in fact, much smaller.
\Cref{fig:pdecay_spectra} shows the total visible energy of the exotic proton decay mode, showing that the energy released into electrons and positrons is significantly smaller than the proton mass.
In this low-energy region, experimental backgrounds, such as from neutrino interactions at underground detectors, become larger, typically by factors of $\mathcal{O}(100)$~\cite{Super-Kamiokande:2020wjk}.
Therefore, constraints on proton decays induced by the muon-induced baryon number violation are expected to be significantly weaker than they are for the golden $p \to e^+\pi^0$ and $p \to e^+e^+e^-$ decay channels.
While a detailed study of the resulting limit is beyond the scope of this paper, it appears that one can safely conclude that the current searches do not exclude lifetimes as small as $10^{32}$~years.

\Cref{fig:pdecay_spectra_momenta} shows the momentum distribution of the final state particles in the exotic proton decay $p\to eee \nu\nu\chi_1\chi_0$.
As expected, the available energy $\Delta E$ is split unevenly between the final states thanks to the large imbalance of masses of the particles involved.
In fact, the final states produced in the chain $\chi_2\to\chi_1 + A^\prime$ carry more energy thanks to the large mass of $\chi_1$ and $\chi_2$.
Similarly, positrons emitted from the $\mu^+ \to e^+ \nu \nu$ chain carry less energy than those from the decay of the on-shell dark photon.

Finally, an important consideration is that the tree-level six-body proton decay channel, for which we estimated the rates above, is not the only possible proton decay mode that can be generated via $\mu p$ annihilation interactions, once radiative corrections are included.
In fact, at the loop level, many new decay channels open up, with a smaller number of final state particles, which relaxes the steep suppression of the decay rate by high powers of $Q$.
An example is a loop correction obtained by attaching the dark photon line to any of the charged particles, which then leads to $p \to (\chi_1 \chi_0) (e^+ \nu_e \overline{\nu_\mu})$ decays.
In similar ways, corresponding decay modes for bound neutrons can be found.
Since these contributions are UV-sensitive, we return to them in \cref{sec:UVcompletions}, where we will show that this reduction in $Q$ dependence is compensated by the necessary inclusion of small couplings/masses.

\subsection{Neutron stars}

Neutron star properties such as mass-radius relation and their cooling function are well understood and can be used to probe many aspects of new physics~\cite{Yakovlev:2004iq,Lattimer:2006xb}. 
Among others, neutron stars can also provide an indirect probe of the new $\mu^-$-nucleon interactions in \cref{eq:general_capture}, due to a large abundance of muons, not only during the initial stage of the core-collapse supernova explosion but also at much later stages in the form of the quasi-degenerate Fermi fluid of $\mu^-$.
A related topic has been addressed in the literature of ``dark neutron" models~\cite{Fornal:2018eol}, from where we borrow several results in our discussion (see, in particular, 
Refs.~\cite{McKeen:2018xwc,Motta:2018rxp,Baym:2018ljz,Cline:2018ami,Strumia:2021ybk,Darini:2022mcy}). 

The zero-temperature nuclear matter inside neutron stars is in chemical equilibrium due to weak interaction transitions, such as $n \leftrightarrow p \ell^- \overline{\nu}_\ell$ and  $p \,\ell^- \leftrightarrow  n \nu_\ell$, where $\ell = \{e, \mu\}$.
These ensure that the chemical potentials $\mu_i$ for $i=n,p,e,\mu,$ satisfy $\mu_n - \mu_p = \mu_e = \mu_\mu$.
Thanks to the large value of $\mu_e$ it is possible for beta-decay type reactions to create muons inside the neutron star despite their larger mass.
The large value of $\mu_e$ also makes $\mu^-$ stable inside the neutron star.
The muon fraction inside the core of the neutron star is determined by nuclear density and equation of state, and, therefore, depends on the modeling of nuclear matter.\footnote{In some models of nuclear matter it is hypothesized that at extremely large densities there is a stable population of heavier baryons and hyperons such as $\Delta(1232)^-$ and $\Sigma^-$, which dominate the number density of negatively charged particle species, and de-leptonize the core.
This can significantly reduce the muon density and, therefore, the rate of $\mu^- p$ or $\mu^- n$ reactions.
Similar arguments apply to the higher-density models where the transition to quark matter reduces the muon chemical potential $\mu_\mu$.}
In Ref.~\cite{Zhang:2020wov}, it was estimated that the fraction of the muon to baryon number densities could be as high as $n_\mu / n_B \sim 20\%$, to be compared with the maximum proton fraction of about $n_p / n_B \sim 40\%$.

The muon-induced aBNV transitions of the form $\mu^- n \to e^- \{X\}^0$ or $\mu^- p \to e^- \{X\}^+$, cf.~\cref{eq:general_capture}, can change the above equilibrium, since now neutrons and/or protons can be converted into dark sector states inside the neutron star.
That is, in the concrete model we are considering in detail,  the process 
$\mu^- p \to \chi_0 \chi_2$ (and the $\nu_\mu n \to \chi_0 \chi_2$ process, generated at 1-loop)
 will become energetically favorable and deplete the number of nucleons.
The capture rate inside the neutron star is given by
\beq
\begin{split}
    \Gamma^{\rm NS}_{\mu^- p \to \chi_0 \chi_2} &= \langle \sigma_{\mu p} v \rangle n_p  \simeq G_{\mu p}^2 m_\mu m_p n_p,
     \end{split}
\eeq
where $\sigma_{\mu p}$ is the cross section for scattering of muons on protons, $v$ is their relative velocity, and $\langle \dots \rangle$ indicates that we should average over the phase space of the particles involved.
For the benchmark value of $G_{\mu p}$,~\cref{eq:benchmarkI,eq:benchmarkII}, this gives
\beq
 \Gamma^{\rm NS}_{\mu^- p \to \chi_0 \chi_2}\simeq \frac{1}{100\,s}\biggr(\frac{G_{\mu p}}{(300\,\text{TeV})^2}\biggr)^2.
\eeq
Interestingly, for dynamics on such intermediate time-scales $\sim \mathcal{O}({\rm min} - {\rm months})$ there are no stringent constraints from the properties of neutron stars, as long as $\chi_0$ and $\chi_2$ do not escape from the stellar interior.
The constraints on properties of the proto-neutron star come from the measurements of the neutrino flux that was generated by the supernova SN-1987a explosion. This probes dynamics on the timescales $\lesssim 10$s of seconds. The constraints on properties of neutron stars, on the other hand, apply only to much longer timescales. 
Typically, there are no direct observations of young neutron stars following the supernova explosion since these are obscured by the remnants of the outer envelope of a progenitor. 

The muon-induced aBNV processes in Eq.~\eqref{eq:general_capture} would heat up the star, 
emitting visible energy in the form of $e^+$ and $e^-$.
The $\mu^-p$ and/or $\mu^-n$ annihilations would create holes in the nucleon and $\mu^-$ Fermi seas, which then get refilled, resulting in additional energy release. 
If such processes are rather slow, {\em i.e.}, occurring on time-scales of millions of years, the late-time observables of neutron stars, such as the temperature~\cite{McKeen:2020oyr,McKeen:2021jbh} of the coldest neutron stars~\cite{Guillot:2019ugf} or the periods of pulsars~\cite{Goldman:2019dbq}, will be sensitive to the muon-induced aBNV.

However, for the capture rates that Mu2e and COMET are sensitive to, the nucleon depletion is fast and will lead to a new equilibrium inside the star with a lower pressure and a softer equation of state (EoS) due to the redistribution of energy over a larger number of degrees of freedom.
The dark sector dynamics will determine whether or not the new EoS will support stars with radii and masses consistent with observation.
These considerations are similar to what happens in mirror neutron models~\cite{McKeen:2018xwc,Motta:2018rxp,Baym:2018ljz,Cline:2018ami}, with the added requirement that muons participate in the conversion reactions. It is clear that the models with stable massive fermions in the dark sector, similar to our benchmark cases, are vastly favored over the models in which the dark sector states decay to very light fermions and/or bosons. 
In the latter case, the newly produced dark states are not retained by the neutron stars and/or do not contribute to the Fermi pressure. 

A new dark force can help restore the pressure lost to the increased number of degrees of freedom.
To illustrate this, consider a scenario where ${\mathcal O}(50\%)$ of nucleons get converted into a population of new fermions, $\chi$.
This may be $\chi_1$ or $\chi_0$, depending on the details of the model.
If $k_{F,0}$ is the original Fermi momentum of neutrons in a neutron star without dark states, then by particle number conservation, $k_{F,0}^3 = k_{F,n}^3 + k_{F,\chi}^3$.
If half of the existing nucleons go to $\chi$, then $k_{F,n} \simeq k_{F,\chi} < k_{F,0}$, thereby reducing the pressure (which is a highly nonlinear function of $k_F$). 

A repulsive self-interaction between the dark sector states contributes to the pressure and could compensate for the reduction of the Fermi momentum. 
The change in pressure due to a single new dark particle $\chi$ subject to a new long-range force mediated by $A^\prime$ is 
\begin{equation}
\Delta P = \frac{1}{15\pi^2}\left(\frac{k_{F,n}^5-k_{F,0}^5}{m_n}+\frac{k_{F,\chi}^5}{m_\chi}\right) + \frac{g_\chi^2k_{F,_\chi }^6}{18\pi^4m_{A'}^2}.
\end{equation}
Therefore, when $(g_\chi^2/\pi^2 m_{A'}^2) \times (k_{F,0} m_n) \gg 1$, this leads to a positive change in pressure compared to the one for a neutron star in the SM,
indicating that it is indeed possible to counteract the loss of pressure through a repulsive force.
In our scenario, the required size of $g_\chi^2/m_{A'}^2$ will depend on the dynamics of the dark sector, the nature of $\chi_0$, and the origin of the new force (the mediator of this dark force may be the dark photon of \cref{sec:darksectormodel}, but in general this may not be the case and could be due to a different light mediator).

An alternative possibility for adjusting the neutron star properties was pointed out in Refs.~\cite{Strumia:2021ybk,Darini:2022mcy}, where the large multiplicity of the dark fermions in the final state was used to correct for the loss of Fermi pressure in the nucleon fluid. In our framework this can be implemented by, {\em e.g.}, further decays along the ``dark decay chain", such as $\chi_1 \to 3 \chi_3$ with $m_{\chi_3}$ not too far from $m_{\chi_1}/3$. This model would then not require any additional pressure from the $A'$ exchanges. 

In conclusion, dark sector models offer sufficient flexibility to compensate for the loss of nucleon Fermi pressure. 
Having outlined a potential reconciliation between observable exotic muon capture and the physics of neutron stars, we leave an investigation of the broader impact of muon-induced baryon number violation on neutron stars to future literature.

\section{UV completions}
\label{sec:UVcompletions}

At first glance, the expected sensitivity of the Mu2e experiment at the level of ${\rm R}(\mu \to e)\sim \mathcal{O}(10^{-16})$ translates to a very impressive NP reach, on the order of $G_{\mu p} \propto 10^{-8} G_F$, or in terms of the effective NP scale, $ G_{\mu p}^{-1/2}\sim 10^3$\,TeV.  
However, $ G_{\mu p}^{-1/2}$ is not a fundamental UV scale. Proton is composed of quarks, and thus we can expect
\beq
\label{eq:Gmup:NDA}
G_{\mu p}\sim \frac{\Lambda_{\rm QCD}^3}{\Lambda_{\rm col}^{d_{\rm col}} \Lambda_{\rm sin}^{5-d_{\rm col}}},
\eeq
where the QCD scale is $\Lambda_{\rm QCD}\sim 0.3$\,GeV, while the value of the exponent, $d_{\rm col}$, depends on the details of the model.  In \cref{eq:Gmup:NDA} we already anticipated that the phenomenologically favored situation is when the mediators charged under QCD are much heavier than the color singlet mediators, and we have thus split the two scales,  $\Lambda_{\rm col}\gg \Lambda_{\rm sin}$ (for simplicity, we still assume that each of the two types of mediators have a common mass).  
 
As we will see below, credible UV models of $\mu^-p$ annihilation to dark states will require \emph{both} heavy new physics, above the electroweak scale, as well as light, GeV scale, states.  
The goal of this section is to demonstrate that such UV completions exist without providing a comprehensive study of all the models that can lead to the effective operator in \cref{eq:modelI}. 

Below, we discuss in detail the triplet-triplet-singlet model that leads to a large scale $\Lambda_{\rm col}$ suppression in \cref{eq:Gmup:NDA} with $d_{\rm col}=3$. We find this to be rather typical, though models with smaller $d_{\rm col}=2$ (and thus a higher value of $\Lambda_{\rm col}$ for which there is still an observable signal at Mu2e and COMET) do exist (see \cref{sec:alternativemodels}).  
It is also important to keep in mind that other UV realizations can lead to distinct LHC and high-energy phenomenology, which can differ significantly from the model we discuss in detail.
 
\begin{figure}
    \centering
    \includegraphics[width=0.47\textwidth]{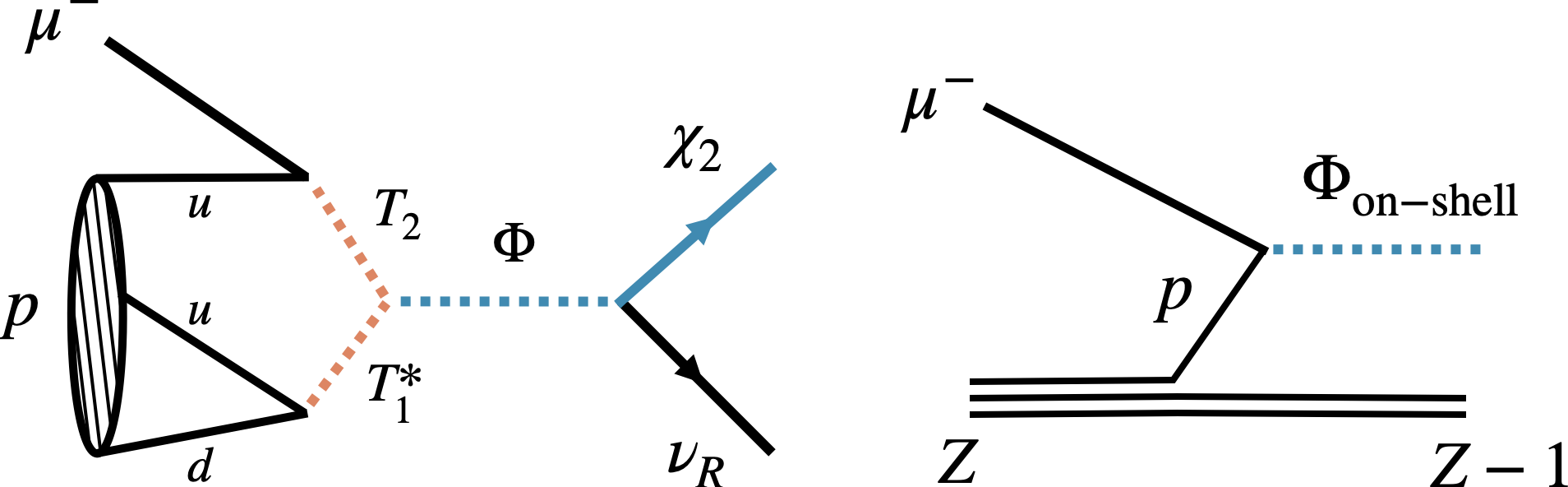}
    \caption{
    The left diagram shows the UV completion of \cref{sec:UVcompletions} that realizes the contact interaction $G_{\mu p}$ of \cref{eq:modelI}. 
    A heavy di-quark $T_1$ and a lepto-quark $T_2$ scalar couple to an intermediate-mass gateway particle $\Phi$.
    The diagram on the right shows how the gateway particle $\Phi$ can be produced on-shell in muon capture with a third-body rescattering ($Z$ represents a nucleus of atomic number $Z$).
    \label{fig:gateway_UV_completions}}
\end{figure}

\subsection{Triplet-triplet-singlet model}
\label{sec:triplet-triplet-singled:model}

A UV model that we consider in detail extends the SM with two scalar color triplets, the di-quark $T_1$ ($Q=-1/3$, $B=-2/3$, $L=0$) and the leptoquark $T_2$  ($Q=-1/3$, $B=1/3$, $L=1$), which have the following couplings to the SM fermions\footnote{For charge conjugate fields we use the notation from \cite{book:Tanagina:Fukugita}. In particular, $\psi_R=P_R\psi=\tfrac{1}{2}(1+\gamma_5)\psi$, and $\psi_R^C=P_L \psi^C=C\bar \psi_R^T=[\psi_R(x)]^C$. (Note that this differs from \cite{Dreiner:2008tw} where $\psi_R^C$ denotes $[\psi_L(x)]^C$.)}
\beq
\label{eq:fullUVscalargateway}
\mathscr{L}  \supset -y_{ud}(\overline{u_{R}^{iC}} d^j_{R}) \epsilon_{ijk} T_1^k 
- y_{\mu u}(\overline{u_R^{iC}} \mu_R) (T_2^*)_i +{\rm h.c.},
\eeq
as well as by a complex color singlet scalar $\Phi$ ($Q=0$, $B=1$, $L=1$), and the dark sector fermions $\chi_0 \equiv \nu_R$ and $\chi_2$ 
\beq\label{eq:gatewayScalar}
\mathscr{L} \supset \rho T_1^{k*} T_{2k} \Phi^* + y_\chi \Phi(\overline{\nu_R}\chi_{2R}^C) +{\rm h.c.}.
\eeq
All the new states are singlets under $SU(2)_L$. The color contractions over $i,j,k$ indices are shown explicitly, while the Lorentz contractions are not; in the latter, $C$ denotes the charge conjugation Dirac matrix. Note that the coupling constants, $y_{ud}, y_{\mu u}, y_\chi$ are dimensionless, while $\rho$ has mass dimension $1$. 
In what follows, we require that $\Phi$ does not develop a vacuum expectation value. If the $\nu_R \chi_2$ bilinear carries the same global quantum numbers as $\Phi$, the above Lagrangian has both $B$ and $L$ conserved. 

We are interested in the limit of very heavy $T_{1,2}$ so that at low energies, these can be integrated out. The resulting effective interaction is
\begin{equation}\label{eq:Phi:quarks}
    \mathscr{L}_{\rm eff}^\Phi=    \frac{1}{\Lambda_{\rm col}^3} \big(\overline{u_R^{iC}} \mu_R\big)
    \epsilon_{ijk}
    \big(\overline{u_R^{jC}} d^k_R\big) \Phi^* +{\rm h.c.},
\end{equation}
where
\beq
\label{eq:Lambda:col}
 \frac{1}{\Lambda_{\rm col}^3} = y_{ud}y_{\mu u}\frac{\rho}{m_{T_1}^2 m_{T_2}^2}.
\eeq
Below the QCD confinement scale, this then leads to the effective interaction with the proton,
\begin{equation}\label{eq:pmuPhi}
    \mathscr{L}_{\rm int}^\Phi = \lambda_{\rm eff} \Phi^{*} \big( \overline{p_R^C} \mu_R\big)+{\rm h.c.},
\end{equation}
with
\beq
 \lambda_{\rm eff} \simeq y_{ud}y_{\mu u}\frac{\rho\,\Lambda^3_{\rm QCD}}{m_{T_1}^2 m_{T_2}^2},
\eeq
where for the matrix element between the tri-quark operator and the proton we used the NDA estimate, 
$\langle 0| \overline{u_{R}^{iC}}(\overline{u_{R}^{jC}} d_R^k) \epsilon_{ijk}| p\rangle\simeq \Lambda_{\rm QCD}^3 \langle 0|\overline{p^C_R}|p\rangle$, with the parenthesis denoting a Lorentz contraction.

\subsubsection{$\Phi$ contributing as an off-shell state}

If $\Phi$ is heavy enough that it can also be integrated out, this gives the effective interaction
\begin{equation}\label{eq:dim9operator}
    \mathscr{L}_{\rm eff} =   \frac{1}{\Lambda_{\rm col}^3}\frac{1}{\Lambda_{\rm sin}^2}  \big(\overline{u_R^{iC}}  \mu_R\big)
    \epsilon_{ijk}
    \big(\overline{u_R^{jC}}d^k_R\big) 
    \left(\overline{\nu_R}\chi_{2R}^C \right)+{\rm h.c.},
\end{equation}
with $\Lambda_{\rm col}$ given in \cref{eq:Lambda:col}, and 
\beq
\label{eq:Lambda:sin}
\frac{1}{\Lambda_{\rm sin}^2}=\frac{y_\chi}{m_\Phi^2},
\eeq
or in terms of interactions with the proton\footnote{The structure of the dimension six operator is similar to the operator considered in \cref{eq:modelI}, but with a different Lorentz structure. This operator further suppresses nucleon decay while leading to ${\mathcal O}(1)$ difference in the muon capture phenomenology. These quantitative differences, however, are not important for the more qualitative discussion we are focusing on here.}
\begin{equation}
\label{eq:pmu:int:dim6}
    \mathscr{L}_{\rm int} =\widetilde G_{\mu p} \left( \overline{p_R^C} \mu_R\right) 
    \left(\overline{\nu_R} \chi_{2R}^C\right),
\end{equation}
where $\widetilde G_{\mu p}$ is given in \cref{eq:Gmup:NDA} with $d_{\rm col}=3$, and $\Lambda_{\rm col}$ and $\Lambda_{\rm sin}$ given in \cref{eq:Lambda:col} and \cref{eq:Lambda:sin}, respectively. 

For observable $\mu^-p$ annihilation rate with $R \simeq 3\times 10^{-15}$, we require, from \cref{eq:R:def}, that $\widetilde G_{\mu p} \simeq 10^{-6} G_F$.  In terms of the UV completion this is,
\beq
\begin{split}
\label{eq:Gmup:estimate}
&\widetilde G_{\mu p}= 
\frac{\lambda_{\rm eff} y_\chi}{m^2_\Phi} 
=2\cdot 10^{-6} G_F \,y_\chi \left(\frac{\lambda_{\rm eff}}{10^{-10}}\right) \left(\frac{2\GeV}{m_\Phi}\right)^2  
\\
&\sim 10^{-6} G_F  \, y_{ud} y_{\mu u}y_\chi
\left(\frac{1 \text{ TeV}}{\sqrt{m_{T_1}m_{T_2}}}\right)^4 
\left(\frac{\rho}{4 \text{ TeV}}\right)\left(\frac{2\text{ GeV}}{m_{\Phi}}\right)^2.
\end{split}
\eeq

In the numerical example in the last line, we show a possible case $m_{T_1,T_2}\sim \rho \gg m_\Phi\gtrsim m_p$.   A light $\Phi$ enhances the $\mu^-p$ annihilation rate, and so would lighter $T_{1,2}$. 
However, the masses of $T_1,T_2$ are required to be at least in the TeV range in order to evade the LHC monojet constraints.
Note that the trilinear scalar coupling $\rho$ was taken to be large.
The $\Phi$ mass receives radiative corrections of size $m_\Phi^2 \sim \rho^2/16 \pi^2$, making our choice of $m_\Phi$ in the GeV range somewhat fine-tuned.

The phenomenology of \cref{sec:darksectormodel} is obtained if $\chi_2$ decays inside the detector to a lighter dark state $\chi_1$ via $\chi_2\to A'\chi_1$, and the dark photon then decays to an electron-positron pair, $A'\to e^+e^-$. We assume that the $\chi_2\to A'\chi_1$ transition is due to a dipole operator $d_{\rm tr}$ in \cref{eq:A':interaction}, and that the states in the SM and in the triplet-triplet-singlet model, including $\chi_{2,1}$, are not charged under the $U(1)_d$. The dipole operator is generated from other heavier states charged under $U(1)_d$ running in the loop. The possibility where $\chi_{2,1}$ are charged under $U(1)_d$ is discussed in \cref{sec:alternativemodels}.

\subsubsection{On-shell $\Phi$ production}

The $\mu^-p$ annihilation rate can be further enhanced if the $\Phi$ scalar is light enough to be produced on-shell, cf.~\cref{fig:gateway_UV_completions}. 
In this case, the muonic atom undergoes a transition $\mu^- (A,Z) \to \Phi (A-1,Z-1)$. The phenomenologically favored kinematics is when the $\Phi$ scalar can be produced on-shell in muon capture, while a proton decay would require both $\Phi$ and $\mu^+$ to be off-shell, thus suppressing the proton decay rate. This occurs for
 $\Phi$ mass in the range $m_p < m_\Phi< m_\mu + m_p$ (ignoring the differences in the nuclear binding energies), while the $\chi_0$ and $\chi_2$ masses satisfy $m_p< m_\mu+m_{\chi_0}+m_{\chi_2}$, with $m_{\chi_0}+m_{\chi_2}<m_\Phi$. 
The observable signal in the $\mu^- p$ annihilation arises from the decay of $\Phi$ to the dark sector states, $\Phi \to \chi_2 \chi_0$, followed by the subsequent decay of $\chi_2$, which involves visible states, such as $\chi_2\to \chi_1 e^+e^-$.

The $2\to 1$ process, $\mu^- p\to \Phi$,  is enabled by the re-scatterings on the other nucleons inside a nucleus. In this sense, the $\mu^- p\to \Phi$ annihilation is analogous to the well-known SM process of pion capture, $\pi^- p \to n$, which at face value is also a $2 \to 1$ process. Following \cref{sec:mucapture_signal} we compute a naive estimate for $\Phi$ production through a ratio with the off-shell channel whose SM-normalized rate is given in \cref{eq:R:def}.
The ratio of the on- and off-shell capture rate is given roughly by 
\begin{equation}
\label{eq:ratio:muZ}
\frac{ \Gamma_{\mu Z \to \Phi (Z-1)}}{ \Gamma_{\mu Z \to \chi_0\chi_2 (Z-1)}}\sim \frac{m_{\Phi,{\mathrm{off}}}^4}{y_\chi^2 E_{\chi_2}E_{\chi_0}}\frac{16\pi^2}{\big(M-m_{\mathrm{Mg}}\big)^3}\tilde r_{\mathrm{p.s.}},
\end{equation}
where the phase space factor can be estimated as
\begin{equation}
\tilde r_{\mathrm{p.s.}}\simeq \frac{\big(1-x_{\Phi,{\mathrm{on}}}^2\big)^{1/2}}{\big(1-x_{02}\big)^2\big(1+x_{02}\big)  }~,
\end{equation}
with $x_{\Phi,{\rm on}}=m_{\Phi,{\mathrm{on}}}/(M-m_{\rm Mg})$ and $x_{02}=(m_{\chi_0}+m_{\chi_2})/(M-m_{\rm Mg})$. 
Here the $m_{\Phi,{\rm on}}$ is the mass of the $\Phi$ that is produced on-shell, while $m_{\Phi,{\rm off}}$ is the mass of the virtual $\Phi$ leading to the three body transition.
Taking $m_{\Phi,\text{on}} = 1 \text{ GeV}$ and comparing to the off-shell benchmark, Eqs.~(\ref{eq:benchmarkI}), (\ref{eq:benchmarkII}), (\ref{eq:Gmup:estimate}), the on-shell rate is significantly enhanced by $\approx 10^9-10^{10} \times y^{-2}_\chi$ depending on the phase space configuration of the three-body off-shell decay. Here, the additional suppression of the three body  off-shell transition, beyond the naive dimensional analysis factor $y_\chi^2/16\pi^2$, comes from the suppression of the matrix element by a factor of $Q \sim E_{\chi_0}$ and the fact that the phase space is more squeezed, resulting in large $\tilde r_{\rm p.s.}$.
The on-shell capture rate is controlled by $\lambda_{\text{eff}}$, where  $\lambda_\text{eff} \lesssim 10^{-14}$ gives capture rates below current $\mu \rightarrow e$ limits, assuming a similar signal of `conversion' electrons to that of the off-shell $\Phi$, see, e.g., Figs.~\ref{fig:muon_capture_spectra} and \ref{fig:muon_capture_spectra_dtr}, after $\Phi$ decays promptly. 
Note that because $R_{\text{on-shell}} \propto \lambda^2_{\text{eff}} \propto (m_{T_1}m_{T_2})^{-4}$ the significant enhancement of the on-shell capture rate only translates to an increase of $m_{T_{1,2}}$ by $\approx$ an order of magnitude.

\subsection{Loop-induced processes}

So far, we have focused on processes directly related, at tree level, to those responsible for the exotic signals at Mu2e.  
Within the triplet-triplet-singlet UV completion, \cref{sec:triplet-triplet-singled:model}, we can also estimate the radiatively generated processes.  
For proton decay, these loop-induced contributions are potentially important since they can lead to proton decay with a smaller number of particles in the final state than those that we used for our estimates in \cref{sec:pdecay}. The radiative corrections can also give rise to new processes, such as bound neutron decays and $n-\chi$ mixing.  
Below, we address the possible constraints due to each of these processes.

\paragraph{Proton decay.}
If any of the final legs in \cref{fig:diagrams} (right) can be contracted to form a loop,\footnote{Remember that we identified $\chi_0=\nu_R$.} this lowers the number of powers of $Q$ suppressing the proton decay rate, and on dimensional grounds also removes the appropriate powers of $G_F$ suppression.  However, contraction of external legs requires small mass and/or coupling insertions.  For instance,  loop-level proton decay $p \to (\chi_1 A^\prime) (\nu_e e^+)$ is obtained by contracting the $\bar \nu_\mu$ and $\nu_R$ external lines, which requires a neutrino mass insertion.  
Contracting  $e^-$ with the $e^+$ line from the weak vertex gives rise to the 5-body decay $p \to e^+ \chi_1 \nu_e \bar{\nu}_\mu \nu_R$, but requires an off-shell $A'$ and is thus suppressed by small kinetic mixing, due to coupling of $A'$ to the $e^+e^-$ pair.
Similarly, the $p \to e^+ \chi_1 \nu_e$ decay, arising from a two-loop diagram with both $\bar \nu_\mu$ contracted with $\nu_R$ and $e^-$ contracted with the $e^+$ line from the weak vertex, is suppressed by the neutrino mass and by kinetic mixing. Because of these suppressions, the loop-induced proton decays always have smaller partial decay widths than the tree-level processes discussed in Sec. \ref{sec:pdecay}.

\paragraph{Neutron decay.}
A one loop $W$ exchange converts the  $u-\mu-T_2$ vertex in \cref{eq:fullUVscalargateway}  to a $d-\nu_\mu-T_2$ one
\begin{align}\label{eq:dressed_u-mu-T_op}
(\overline{u^{iC}_R} \mu_R) (T_2^*)_i \rightarrow  \frac{y_\mu y_u}{16\pi^2}  \times (\overline{d_L^{iC}} \nu_{\mu} ) (T_2^*)_i~,
\end{align}
which after the $T_{1,2}, \Phi$ are integrated out, and the higher dimension operator is run down to the weak scale, gives the effective Lagrangian 
\begin{equation}
\label{eq:nmu:int:dim6}
    \mathscr{L}_{\rm int} =\kappa \widetilde G_{\mu p} \left( \overline{n_L^C} \nu_\mu\right) 
    \left(\overline{\nu_R} \chi_{2R}^C\right),
\end{equation}
where
\begin{equation}\label{eq:kappa}
    \kappa \simeq \frac{y_\mu y_u}{16\pi^2} \log\left(\frac{\Lambda_{\rm UV}^2}{M^2_W}\right) \sim 10^{-10}~,
\end{equation}
with $\Lambda_{\rm UV}\sim m_{T_1}, m_{T_2}$.
The muon and up-quark Yukawa couplings arise due to mass insertions required for converting right-handed fields to left-handed ones, which then couple to the $W$ boson. 

In the same way as the $u-\mu-T_2$ vertex leads to the proton decay inducing operator in \cref{eq:dim9operator}, after $T_{1,2}$ and $\Phi$ are integrated out, the $d-\nu_\mu-T_2$ interaction in \cref{eq:dressed_u-mu-T_op} will induce the decay of the neutron into dark sector particles, 
\begin{equation}\label{eq:neutron_decay}
    n \to (\chi_2^*\to \chi_1 A^\prime) \nu_R \bar\nu_\mu~,
\end{equation}
however, with a highly suppressed rate proportional to $\kappa^2$, 
\begin{equation}\label{eq:neutron_lifetime}
\Gamma_n \simeq \left(\frac{g_D \kappa\, G_{\mu p}}{m_2}\right)^2  \frac{Q^7}{\left(16\pi^2\right)^2}.
\end{equation}
Here, $Q\approx (m_n-\epsilon_b-m_1-m_{A'}-m_{\nu_R})/4$ is the typical momentum of the final state particles. 
In the expression for it, we already anticipated that the phenomenologically most relevant are the decays of neutrons bound inside nuclei, with $\epsilon_b$ the average binding energy of a neutron inside a nucleus.
This binding energy is large enough that the bound neutron decay will be kinematically forbidden for the squeezed mass spectra we consider.

The bounds from searches for exotic decays of free neutrons of the type $n\rightarrow e^+e^-+\mathrm{inv}$ are relatively weak \cite{UCNA:2018hup,Klopf:2019afh}.  
However, if these transitions occur inside a nucleus, they would appear as decays of stable or long-lived nuclei, which then places stringent constraints on new physics models. 
In particular, $^{12}$C$\to ^{11}$C$ + \mathrm{inv}$ and $^{16}$O$\to ^{15}$O$+ \mathrm{inv}$ decays are well constrained, since the de-excitations of the daughter nucleus would leave visible signals in the underground large-volume detectors such as Borexino~\cite{Borexino:2003igu}, KamLAND~\cite{KamLAND:2005pen}, and SNO/SNO+~\cite{SNO:2003lol,SNO:2022trz}. 
Since the daughter nuclei are meta-stable, the de-excitation signatures can be targeted experimentally, independently of the electron/positron pair signature that arises from the decay of $A^\prime$.  
A lack of any positive signal places a bound $\tau_n\gsim 10^{29}$ years, which should be compared with the expectations in \cref{eq:neutron_lifetime}, shown in the right panel of \cref{fig:plifetimes}. 
Note that we take $\kappa\sim 10^{-10}$ and have neglected the additional penalty from the binding energy.

In summary, the operator behind neutron decay indeed lifts part of the phase-space suppression in free proton decay but also gets suppressed in our model by $u$-quark and muon mass insertions. Due to the binding energy of the nucleus, it may even be kinematically forbidden in part of the parameter space.

\begin{figure}[t]
    \centering
    \includegraphics[width=0.8\columnwidth]{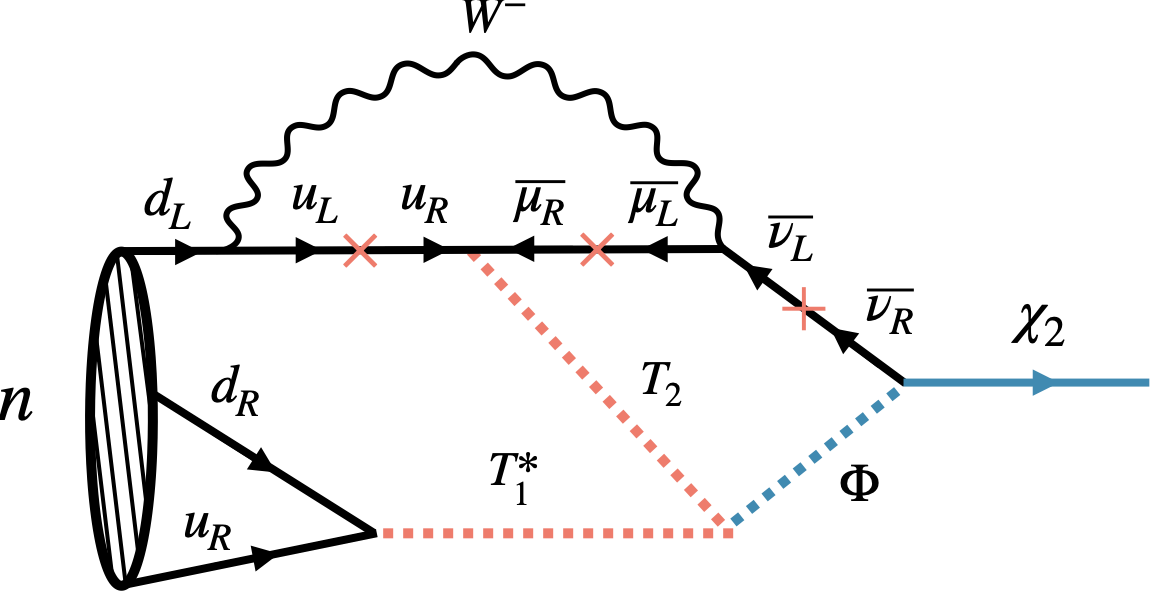}
    \caption{Neutron mixing with $\chi_2$ generated at the two-loop level. 
    At an additional loop, the mixing between $\chi_1$ and $n$ is generated by an $A^\prime$ line.
    \label{fig:n-chi-mixing} }
\end{figure}

\paragraph{$n-\chi$ mixing.}
Since in our specific UV completion $\chi_0 = \nu_R$, the diagrams such as the one in \cref{fig:n-chi-mixing} induce mixing between $n$ and $\chi_{2}$ ($\chi_1$) at two-loops (three-loops), giving rise to $n \to \chi_1$ transitions.
The relevant operators are
\begin{equation}\label{eq:nchi_mixing}
\mathscr{L}_{\rm eff}\supset m_{n\chi_1} \bar  n \chi_{1} + m_{n\chi_2} \bar  n \chi_{2} + \text{h.c}.
\end{equation}
The $m_{n\chi_1}$ parameter has greater phenomenological significance as it leads to kinematically-allowed $n\to\chi_1$ transitions in nuclei.
Let us first estimate $m_{n\chi_2}$ in our UV completion using naive dimensional analysis,
\begin{align}\label{eq:nchi_mixing_estimates}
    m_{n \chi_2} &\sim \frac{y_u y_\mu y_\nu}{(16 \pi^2)^2} (\lambda_{\rm eff} y_\chi) v_{\rm EW}\log^2\left(\frac{\Lambda_{\rm UV}^2}{m_{W}^2}\right)
    \\\nonumber 
    &\sim y_\chi \times 10^{-28} \text{ MeV} \times \left(\frac{\lambda_{\rm eff}}{10^{-10}}\right)\left(\frac{y_\nu}{10^{-12}}\right).
\end{align}
Note that the mixing parameter is proportional to the neutrino coupling $\kappa$ in \cref{eq:kappa} and neutrino masses.
One more loop is then required to transform $\chi_2$ into $\chi_1$ and is generated by an $A^\prime$ line that either connects to $\nu_R$, if charged under $U(1)_D$, or to an electrically charged state through kinetic mixing.
We discuss the possible $U(1)_D$ charge assignments in \cref{sec:alternativemodels}.  
For now, we focus on the latter possibility, which leads to an estimate of $m_{n \chi_1} \sim m_{n \chi_2} \times \epsilon^2\alpha_D/4\pi$.

The rate for neutron disappearance inside a nucleus is given by Fermi's golden rule,
\begin{equation}
\Gamma_n = 2\pi \left| \langle \psi_f|m_{n\chi_1} |\psi_i\rangle\right|^2 \delta(E_f-E_i) \rho(E_f)~,
\end{equation}
with $\rho(E_f)$ the density of states of the final state and $\psi_i$ and $\psi_f$ are the wavefunctions of the initial neutron and final state $\chi_1$ state, respectively.  
The momentum of the final state, a free $\chi_1$ particle, is determined by energy conservation with $p_f^2/2m_1=m_n-\epsilon_b-m_1$ and $\epsilon_b$ is the binding energy of the neutron. 
Taking into account that there are $N=A-Z$ neutrons inside a nucleus, the rate for nuclear transition induced by the $n-\chi_1$ mixing is given by
\begin{equation}\label{eq:nchinuclearrate}
\Gamma_{\mathrm{N}} = \frac{N}{\pi} m_{n \chi_1}^2 m_1 p_f \xi^2 V_{\mathrm{nucl.}}~.
\end{equation}
The overlap of the final state wavefunction with the neutron, $\psi_i$, bound inside a nucleus of volume $V_{\mathrm{nucl.}}$, is encoded in the dimensionless quantity $\xi$, where
\begin{equation}\label{eq:wavefunctionoverlap}
\xi^2 = \left|\frac{1}{\sqrt{V_{\mathrm{nucl.}}}} \int d^3 r\, \psi_i \exp\left(i \vec{p} \cdot \vec{r}\right) \right|^2~.
\end{equation}
For large energy release, $p_f^3 \gg V_{\mathrm{nucl.}}$, this overlap is small.  Instead, we assume $\xi\approx 1$ and arrive at an estimate for the decay time of a carbon nucleus, where $\epsilon_b \sim 10\MeV$, due to neutron disappearance of
\begin{equation}\label{eq:neutron-disappear-estimate}
\begin{split}
\tau_{\mathrm{C}} \simeq\, & 10^{33}\,\mathrm{years}\times \left(\frac{850\MeV}{m_1}\right)\left(\frac{100\MeV}{p_f}\right) \\
&
\times \left(\frac{100\,\mathrm{fm}^3}{V_{\mathrm{nucl.}}}\right)\left(\frac{10^{-28}\MeV}{m_{n \chi_1}}\right)^2~.
\end{split}
\end{equation}
This lifetime should be compared to bounds on Pauli exclusion transitions from Borexino \cite{Borexino:2004hfc}, which searched for a spontaneous formation of holes in the closed nucleon shells of $^{12}$C.  
For both the Pauli excluded transitions or a neutron conversion to $\chi_1$ in a carbon nucleus, the remaining nucleons will refill shells, releasing an $\mathcal{O}(10\,\MeV)$ photon.  
Borexino's lack of a signal places a constraint of $\tau_{\mathrm{C}} \gsim 10^{27}$ years, which is easily satisfied in our model.

\subsection{Alternative scenarios}
\label{sec:alternativemodels}

Above, we discussed only one possible UV completion in detail, the triplet-triplet-singlet model. While this model is a good representative of how the experimental constraints can be evaded, primarily though compressed spectra, we expect that many other SM extensions exist that can lead to observable $\mu-p$ annihilation while satisfying current constraints. Below, we discuss some of the possible variants of the triplet-triplet-singlet model as well as its extensions. Some of these variants have already been touched upon above. 

\paragraph{Alternative scenarios for $\Phi$ produced on-shell.}
In the numerical analysis, we primarily focused on the case of a heavy singlet $\Phi$. However, the $\Phi$ can also be lighter, even so light that it is produced on-shell in $\mu^-p$ annihilation. The $\Phi$ can then decay into the dark sector states and the $e^+e^-$ pair, which is the same final state particle content as the one we discussed in detail in the case of heavy $\Phi$, but with modified kinematics. It is also possible that the dominant decay of $\Phi$ is into other visible particles (and dark states), not the $e^+e^-$ pair. For instance, $\Phi\to 4e+$dark states or $\Phi\to \gamma+$dark states are both possible and can have drastically different experimental efficiencies at an experiment such as Mu2e.

\paragraph{Production of kaons.}
An interesting alternative for releasing more energy is $\mu^-p$ annihilation to kaons. This could, for instance, be mediated via the following low-energy interactions, $\mathscr{L} \supset \lambda_{\rm eff} \Phi^* \overline{p_R^C} \mu_R + \lambda_{KK} \Phi (K^0)^2$.
For $\Phi$ that has a mass in the range $2m_K < m_\Phi < m_p + m_\mu$, the muon induced conversion process is then $\mu^- p \to \Phi +{\rm nucl\, recoil} \to K^0K^0$. Proton decay, on the other hand, can only proceed via two virtual particles, $p \to (\mu^+)^* (\Phi^* \to K K^*)$.
As before, the main idea is to use the muon mass as a means to enable baryon number violation in $\mu p$ capture while at the same time suppressing proton decay via a judicious choice of the mass spectrum. 
The novelty in this scenario is that the byproducts of the capture reaction are fixed by the branching ratios of the kaon, leading to the production of pions, muons, and photons in addition to electrons and positrons.
Note, however, that this scenario is quite likely challenged by the neutron star physics. 
In the neutron star, the final state $K$-mesons will get re-absorbed by nuclei, resulting in a loss of two fermions per every $\mu^-p$ annihilation, with strong implications for the properties of the neutron stars.

\paragraph{Alternative dark photon couplings.} 
In the main part of the text, we assumed that the $\chi_2\to \chi_1A'$ transition proceeds through dimension 5 dipole operators. However, this transition can also be mediated by renormalizable interactions. To show this, we first assume that the dark photon mass arises from its coupling to dark higgs, $h_d$, after it acquires a vev. The interactions of the dark $U(1)_D$ gauge boson, $A_\mu^\prime$, depend on the assigned $U(1)_D$ charges of the dark states $\chi_{1,2}, \nu_R$.  To allow for the transition $\chi_2\rightarrow \chi_1 A'$, $\chi_1$ and $\chi_2$ must have the same charge.  To avoid charging SM particles under the $U(1)_D$, $\nu_R$ must be oppositely charged to $\chi_{1,2}$, or the second term in \cref{eq:gatewayScalar} must contain an insertion of $h_d$.  One possibility is that none of these carry a $U(1)_D$ charge, and thus the $\chi_2\to \chi_1 A'$ decay is due to the dipole interaction in \cref{eq:A':interaction}. If the $\chi_i$'s carry a $U(1)_D$ charge, the decay takes place through the contact operator in \cref{eq:A':interaction}.  
In either case, the coupling to the dark photon is off-diagonal in $\chi$ flavor space.  

This off-diagonal form of the coupling can be achieved if there are 3 Weyl fermions in the dark sector, two of which make up a pseudo-Dirac pair.  
Consider 3 states $\psi_0, \psi_1, \psi_1^c$, with a vector-like mass $m\psi_1 \psi_1^c$ and a small Majorana mass $\delta \psi_1 \psi_1$.  This can be generated, for instance, by integrating out a massive fermion with a small Yukawa coupling with $\psi_1$ and $h_d$.  
The mass eigenstates $\chi_{1,2}$ are approximately equal admixtures of $\psi_1$, $\psi_1^c$. Since Majorana fermions can neither carry a charge nor have a magnetic dipole moment, the mass mixing implies that the couplings will be of the transition type.  
If $\psi_i$ carry a magnetic dipole moment\footnote{These dipole moments may be generated perturbatively through loops of $U(1)_D$ charged states, in which case the typical size is $\sim g_D^3/(16\pi^2 m_\pm)$. 
Or they may be generated non-perturbatively, as for the SM baryons, in which case we expect the size to be $\sim g_D/m_\chi$.} the mass eigenstates $\chi_{1,2}$ will acquire a magnetic dipole transition operator.  If instead $(\psi_0, \psi_1, \psi_1^c)$ carry charge $(1,1,-1)$ respectively, with the massive state which generates the Majorana mass having charge 2, then $\chi_{1,2}$ will couple in an off-diagonal fashion to the dark photon.

Even if the SM fields are not charged under the dark gauge group, the coupling of the dark photon to the SM fields can occur through kinetic mixing of $U(1)_D$ with $U(1)_Y$. This would be generated at one loop by having fields charged under both groups running in the loop.  At low energies, this leads to a kinetic mixing between the dark photon and the SM photon of size $\epsilon\sim e g_D/(16\pi^2)$.  
The dark photon couplings to charged leptons are then of size $\epsilon e$.

\paragraph{Models that lead to $\mu^-n$ annihilation.}
A completely separate set of models that can lead to energetic electrons at Mu2e and COMET are models that have the $\mu^-n$ annihilation as the leading effect, while $\mu^-p$ is induced only at the subleading loop level. An example are models that result in
neutron-to-dark-state $\chi_n$ transition, induced by the mass mixing $\mathscr{L}\supset \epsilon (\bar{n}\chi_n + {\bar\chi_n}n)$, as in Ref.~\cite{Fornal:2018eol}.
If the UV model contains $\mu\to e$ CLFV interactions, then the $\mu^- n \to e^- \chi_2$ transition becomes possible. 

\section{Conclusions}
\label{sec:conclusions}

In this manuscript, we have constructed new physics scenarios with muon-induced (apparent) baryon number violation.
Using a specific example, we demonstrated that these models lead to a spectrum of higher energy electrons and positrons in the $\mu^- A \to e^- A$ conversion experiments, such as Mu2e and COMET, while satisfying experimental bounds on proton stability.
The main idea consists of tapping into the baryonic energy reservoir of the nucleus $A$ by 
destroying protons or neutrons in the muon capture process.
This releases energies greater than $m_\mu$ into dark particles that eventually decay to $e^+e^-$ pairs.
The $e^+$ or $e^-$ can reach energies as high as $\mathcal{O}(130)$~MeV, depending on the choice of parameters, significantly overshooting the energy of electrons produced in decay-in-orbit as well as of eventual monochromatic
electrons from $\mu^- \isotope[27]{Al} \to e^- \isotope[27]{Al}$ ($E_{e^-}^{\rm Al-conv} = 104.98$~MeV). 

These signatures are not excluded by limits on the proton lifetime thanks to a coincidence between the mass of the proton and the dark particles, which forces the protons to decay via two or more off-shell states.
The stability of nuclei (more specifically, that of bound neutrons) also imposes meaningful constraints due to radiative corrections.
These are again bypassed due to strong phase-space suppression as well as by chiral suppression in the UV completions we considered. The crucial component for constructing models that enhance the $\mu^--$nucleon annihilation while suppressing nucleon decays is that the ratio of muon to proton mass is $m_\mu/m_p\sim 1/9$, which is not a very small ratio. 
Consequently, changes in mass splittings in the dark sector at the level of $\sim 20$\,MeV ({\em i.e.}, ~at a few percent level) are very consequential for the signals discussed here and can change answers by many orders of magnitude. 

While our scenarios are specifically designed to hide nucleon decay and highlight muon capture, they demonstrate that Mu2e and COMET can be sensitive to electrons that come from new physics sources other than the $\mu \to e$ lepton flavor violating neutral currents.
Such scenarios are particularly attractive in the event that high-energy electrons or positrons are detected at Mu2e.
Our proposal also shows that such high-energy events may not necessarily come only from backgrounds such as the decay of late-time pions or cosmics.

Finally, we also emphasize that positrons may present a particularly interesting final state.
Light dark sectors can lead to positron energies beyond those induced by radiative muon capture and by the lepton-number-violating channel $\mu^- A\to e^+ A'{}^*$~\cite{Geib:2016atx,Berryman:2016slh}.
The latter includes a monochromatic positron with energy of about $E_{e^+}^{\rm conv} \simeq 92.3$~MeV for $\mu^- \, \isotope[27]{Al} \to e^+ \, \isotope[27]{Na} ({\rm g.s.})$ transitions.
It should be noted that past searches for $\mu^- \to e^+$ at TRIUMF~\cite{Ahmad:1988ur} and PSI~\cite{SINDRUMII:1993gxf,SINDRUMII:1998mwd,SINDRUMII:2006dvw} on Ti and Au targets have all reported excesses of positrons above $E_{e^+} \gtrsim 90$~MeV but below $E_{e^+} \lesssim m_\mu$ (see also earlier searches in \cite{Conforto:1962zz,Bryman:1972rf,Abela:1980rs}).
While the modeling of radiative muon capture may be insufficient in this energy region~\cite{Lee:2021hnx}, these events represent an immediate application of our scenarios.
In fact, since the positron energies in these excess events are still less energetic than $m_\mu$, an explanation through muon-induced BNV would be feasible even for parameters that render protons and bound neutrons stable.

\section*{Acknowledgements} 
We want to thank David McKeen for valuable discussions at the Muons in Minneapolis workshop 2023, and Bertrand Echenard during the 2024 summer program at Aspen Center for Physics.
We acknowledge support from the Simons Foundation Targeted Grant 920184 to the Fine Theoretical Physics Institute. 
JZ and TM  acknowledge support in part by the DOE grant de-sc0011784 and NSF OAC-2103889. TM acknowledges support in part from the Visiting Scholars Award Program of the Universities Research Association. MP is supported in part by the DOE grant DE-SC0011842.
The work of MH is supported by the Neutrino Theory Network Program Grant \#DE-AC02-07CHI11359 and the US DOE Award \#DE-SC0020250.  PJF is supported by the DOE under contract \#DE-SC0007859 and Fermilab, operated by Fermi Research Alliance, LLC under contract \#DE-AC02-07CH11359 with the United States Department of Energy. This work was performed in part at the Aspen Center for Physics, which is supported by National Science Foundation grant PHY-2210452.

\bibliographystyle{JHEP}
\bibliography{muon_induced_BNV}
  
\end{document}